\documentclass[11pt]{article}
\usepackage{eurosym}
\usepackage{amsfonts}
\usepackage{amssymb}
\usepackage{graphicx}
\usepackage{amsmath}
\usepackage{makeidx}
\usepackage{indentfirst}
\usepackage[T1]{fontenc}
\usepackage[utf8]{inputenc}

\newcounter{resultnum}[section]
\newcounter{conclusionnum}[section]
\newcounter{conditionnum}[section]
\newcounter{conjecturenum}[section]
\newcounter{examplenum}[section]
\newcounter{exercisenum}[section]
\newcounter{lemmanum}[section]
\newcounter{notationnum}[section]
\newcounter{theoremnum}[section]
\newcounter{definitionnum}[section]
\newcounter{corollarynum}[section]
\newcounter{remarknum}[section]
\newcounter{propositionnum}[section]
\newcounter{acknowledgementnum}[section]
\newcounter{algorithmnum}[section]
\newcounter{axiomnum}[section]
\newcounter{casenum}[section]
\newcounter{claimnum}[section]
\newcounter{summarynum}[section]
\newcounter{problemnum}[section]
\begin{document}

\title{Nonassociative cosmological solitonic R-flux deformations  in \\ gauge gravity and G. Perelman geometric flow thermodynamics}
\date{April 2, 2024}
\author{ {\textbf{Lauren\c{t}iu Bubuianu}\thanks{%
email: laurentiu.bubuianu@tvr.ro and laurfb@gmail.com}} \and {\small \textit{%
SRTV - Studioul TVR Ia\c{s}i} and \textit{University Appolonia}, 2 Muzicii
street, Ia\c{s}i, 700399, Romania} \vspace{.1 in} \\ 
{\textbf{Julia O. Seti }} 
\thanks{email: j.seti@chnu.edu.ua}\\ 
{\small \textit{\ Department of Information Technologies and Computer Physics }}\\
{\small \textit{\ Yu. Fedkovych Chernivtsi National University, Chernivtsi, 58012, Ukraine; }}\\ 
{\small \textit{\ Department of Applied Mathematics, Lviv Polytechnic National University, }}\\
{\small \textit{\ Stepan Bandera street, 12, Lviv, 79000, Ukraine}}  \vspace{.1 in} \\
\textbf{Sergiu I. Vacaru} \thanks{%
emails: sergiu.vacaru@fulbrightmail.org ; sergiu.vacaru@gmail.com }  
\\
{\small \textit{\ Volunteer researcher in Chernivtsi, Ukraine }}
 \vspace{.1 in} 
\and {\textbf{El\c{s}en Veli Veliev }} \thanks{email: elsen@kocaeli.edu.tr and elsenveli@hotmail.com} \\
{\small \textit{\ Department of Physics,\ Kocaeli University, 41380, Izmit, Turkey }}
\vspace{.1 in}
}
\maketitle

\begin{abstract}
We elaborate on a model of nonassociative and noncommutative gauge gravity for the de Sitter gauge group $SO(4,1)$ embedding extensions of the affine structure group $Af(4,1)$ and the Poincar\'{e} group $ISO(3,1)$. Such nonassociative gauge gravity theories are determined by star product R-flux deformations in string theory and can be considered as new avenues to quantum gravity and geometric and quantum information theories. We analyze physically important and geometric thermodynamic properties of new classes of generic off-diagonal cosmological solitonic solutions encoding nonassociative effective sources. Particularly, we focus on modelling by such solutions of locally anisotropic and inhomogeneous dark matter and dark energy structures generated as nonassociative solitonic hierarchies. Such accelerating cosmological evolution scenarios can't be described in the framework of the Bekenstein-Hawking thermodynamic formalism. This motivates a change in the gravitational thermodynamic paradigm by considering nonassociative and relativistic generalizations of the concept of W-entropy in the theory of Ricci flows. Finally, we compute the corresponding modified G. Perelman's thermodynamic variables and analyze the temperature-like evolution of cosmological constants determined by nonassociative cosmological flows.

\end{abstract}

\newpage

\tableofcontents


\section{Introduction and preliminaries}

\label{sec1}

The nonassociative and noncommutative geometry offers intriguing possibilities and new insights to study the structure of fundamental interactions with links to string and modified  gravity theories and generalizations of the standard model of particle physics. A class of such theories is elaborated for star products, $\star $, determined by so-called twisted R-flux deformations when generalizations of the geometric and quantum field notions result in new type models of classical and quantum spacetime. The nonassociative $\star $-formalism can be employed to investigate various aspects of black holes and cosmology physics based on non-Riemannian geometry and modified gravity theories, MGTs. In \cite{blumenhagen16,aschieri17}, such physically viable nonassociative MGTs were constructed up to the definition of nonassociative vacuum Einstein equations, $\mathcal{R}ic^{\star }[\mathbf{g}^{\star },\mathbf{\nabla }^{\star }]=0,$ for a corresponding star product deformed variant of the Ricci tensor $\mathcal{R}ic[\mathbf{\nabla }]$ of the Levi-Civita, LC, connection $\mathbf{\nabla }$.\footnote{In this paper, abstract star labels are used for denoting star product deformations, for instance, of (pseudo) Riemanian metrics, $\star :\mathbf{%
g\rightarrow g}^{\star }$; of the LC-connection, \ $\star :\ \nabla \rightarrow \nabla ^{\star }$; and other nonassociative $\star $-variants of fundamental geometric objects like Riemannian curvature tensor, Ricci tensor  $\mathcal{R}ic^{\star }$, etc. In explicit form, the components of geometric objects and their coefficients can be computed for parametric decompositions on the string constant $\kappa $ and Plank constant $\hbar .$ Appendix \ref{appendixa} contains a summary of definitions and formulas on nonassociative star product geometry.} Then, the approach was extended in abstract \cite{misner} and nonholonomic geometric form \cite{partner02,partner04} for nonassociative gravity theories with effective nontrivial sources by considering modifications of general relativity, GR. Such modified gravity theories, MGTs, involve twisted scalar products adapted to nonholonomic distributions on nonassociative phase spaces $\ _{\star }\mathcal{M}$ defined as star product deformations of Lorentz manifolds on cotangent
Lorentz bundles enabled with nonholonomic star products.

\vskip4pt In the search for the unification of GR and standard particle physics, many attempts have been made to formulate and study different models of gauge gravity and related MGTs. Since the first papers 
\cite{utiyama,kibble} there were elaborated hundreds of such models. We omit to present a complete historical account but refer to several available reviews on (non) commutative geometry and gravity: \cite{hehl,jurco,svnc00,vd00}, for early works, and \cite{svmp05,ciric}, for a method of constructing
gravitational diagonal or off-diagonal solutions on noncommutative phase spaces.\footnote{%
In this letter, we shall elaborate on a variant of nonassociative star product deformation of GR using the de Sitter gauge group $SO(4,1)$ on $\ _{\star }\mathcal{M},$ but the constructions can be performed similarly, for instance, using $SO(2,3)_{\star }$ and other types of base (non) commutative manifolds.}

\vskip4pt Efforts are currently in progress for using MGTs to explain modern observational cosmological data \cite{foroconi23,boylan23,biagetti23} and study possible applications to, dark energy, DE, and dark, DM, physics \cite{saridakis,mavromatos,antoniadis} (we mention just a few of them). The main motivation to elaborate on nonassociative and noncommutative geometric and physical models stems from string/M-theory and/or supersymmetric Yang-Mills, YM, theories \cite{seiberg,ferrari,condeescu,blumenhagen13}. In the context of string gravity, such approaches were typically based on constructions using the Seiberg-Witten map (SW, and noncommutative star product) with (non) linear gauging of certain structure groups and/or supersymmetric/ quantum generalizations \cite{jurco,svnc00,sv00}. In the approach based on noncommutative geometry, many works have been published, for instance, in the context of black holes, BHs, physics and the study of their evaporating and thermodynamic properties \cite{svmp05,chamseddine,nicolini,chaichian}.
Nevertheless, up to now, there are gaps in the literature concerning nonassociative gauge gravity theories, in constructing nonassociative cosmological solutions and investigating DE and DM thermodynamic effects
which may encode mixed nonassociative and noncommutative structures.

\vskip4pt This work has three main goals stated respectively for each section:

\begin{enumerate}
\item The purpose of section \ref{sec2} is to formulate in geometric form a nonassociative gauge model version of nonassociative gravity theories determined by star product deformations as in \cite{blumenhagen16,aschieri17} and (in nonholonomic form, which is crucial for applying the anholonomic frame and connection deformation method, AFCDM, for constructing exact and
parametric solutions) \cite{partner02,partner04}. Such a nonassociative gravity theory is a nonassociative and noncommutative phase space $\ _{\star}\mathcal{M}$ generalization of the noncommutative and nonholonomic constructions from \cite{svnc00,vd00,sv00}. The nonassociative star product deformed YM-type equations (i.e. gravitational equations) will be derived geometrically in such a way that in the commutative limit, they result in standard Einstein equations for GR, when nonassociative contributions can be computed for effective sources with decompositions on constant parameters $\kappa $ and $\hbar .$

\item The second purpose, in section \ref{sec3}, is to show that by applying the AFCDM we can construct cosmological solutions, which in terms of metric components written in coordinate bases are generic off-diagonal and may depend on spacetime and phase space coordinates. Such solutions, see
subsection \ref{ssec31}, are determined by solitonic hierarchies describing off-diagonal solitonic wave cosmological evolution scenarios \cite{sv15a} and characterised by nonlinear symmetries (gauge type, Killing ones etc.) relating effective sources of matter encoding nonassociative R-flux data from string theory to certain effective cosmological constants \cite{partner02,partner04}.

\item Then, the third purpose (for subsection \ref{ssec32}) is to speculate how new types of off-diagonal cosmological solutions in nonassociative gauge gravity can be applied for describing certain models and nonlinear effects for DE (defined by running of effective cosmological constants) and DM (determined by generating solitonic functions and effective generating sources). We shall argue that such nonassociative cosmological models can't be described in the framework of the Bekenstein-Hawking paradigm 
    \cite{bek2,haw2} but corresponding geometric/ statistical thermodynamic models can be elaborated using G. Perelman's concept of W-entropy \cite{perelman1}. Such constructions can be performed for (non) commutative and nonassociative phase spaces \cite{svnonh08,partner04}, see also \cite{kehagias19} on recent applications of the theory of Ricci flows in modern high-energy physics.
\end{enumerate}

Finally, in section \ref{sec4}, we briefly conclude the results and discuss further perspectives for nonassociative gauge gravity theories. Appendix \ref{appendixa} contains necessary definitions and formulas on nonassociative star product geometry (readers are recommended to familiarise themselves with those results and \cite{partner02} before studying the next sections of this paper). In Appendix \ref{appendixb}, we outline the main ideas and formulas on nonlinear solitonic hierarchies and cosmological solitonic waves.

\section{Nonassociative star product deformations of affine and de Sitter gauge gravity}

\label{sec2}

In this section, we formulate a model of nonassociative gauge theory of gravity generalizing the (non) commutative gauge gravity theories from \cite{svnc00,vd00,sv00}. Geometric tools to address nonassociative gravity theories with twister star product are provided in Appendix \ref{appendixa} as a summary of necessary results from \cite{partner02,partner04} and references therein.

\subsection{A nonholonomic commutative model of gauge phase space gravity}

In this paper, the commutative geometric arenas consist from phase spaces $%
\mathcal{M}=T\mathbf{V}$ and $\ ^{\shortmid }\mathcal{M}=T^{\ast }\mathbf{V}$
modelled repectively as relativistic tangent and cotangent bundles. A base
(conventional, horizontal, h) spacetime Lorentz manifold $\mathbf{V}$ can be
considered as in GR , when the (horizonatl) metric $hg$ is of local Minkowski signature
(+++-). A total metric $\ ^{\shortmid }\mathbf{g}$ for $\ ^{\shortmid }%
\mathcal{M}$ is defined as symmetric second rank tensor which can be written
in abstract form (following a generalized index/coordinate free formalism
developing that from \cite{misner,partner02,partner04}) as a distinguished
metric, d-metric, $\ ^{\shortmid }\mathbf{g=}(h\ ^{\shortmid }g=hg,c\
^{\shortmid }g),$ where $c\ ^{\shortmid }g$ is stated for a dual/ covector
typical co-fiber. A total metric $\mathbf{g}$ for $\mathcal{M}$ can be
introduced for $h$ and $v$-splitting by a nonlinear connection,
N-connection, $\mathbf{N}$, or for a shell nonholonomic dyadic decomposition
(including dual phase spaces) as stated by formulas in Appendix \ref%
{appendixa}. To elaborate gauge theories on $\mathcal{M}$ or $\ ^{\shortmid }%
\mathcal{M},$ we shall use a gauge structure groups $\mathcal{G}r=\left(
SO(4,1),SO(4,1\right) )$ where the de Sitter gauge group $SO(4,1)$ encodes 
consequent nonlinear extensions of the affine structure group $Af(4,1)$ and
the Poincar\'{e} group $ISO(3,1)$ as it is physically motivated in \cite%
{jurco,svnc00,sv00,vd00}. Then, we assume that the commutative geometric
parts of our models will involve commutative nonholonomic (co) vector bundle
spaces 
\begin{equation}
\ ^{\shortmid }\mathcal{E(\ ^{\shortmid }M)}:=\left( \ ^{\shortmid }\mathcal{%
E}=h\ ^{\shortmid }\mathcal{E}\oplus c\ ^{\shortmid }\mathcal{E},\
^{\shortmid }\mathcal{G}r=SO(4,1)\oplus \ ^{\shortmid }SO(4,1),\ ^{\shortmid
}\pi =(h\pi ,c\pi ),\mathcal{\ ^{\shortmid }M}\right) ,  \label{dvbundle}
\end{equation}%
associated to respective tangent bundles $T\mathcal{(\ ^{\shortmid }M)}$ and
co-tangent bundles $T^{\ast }\mathcal{(\ ^{\shortmid }M)}$ enabled with
N-adapted projections $\pi $ and $\ ^{\shortmid }\pi .$ In brief, we shall
write $\ ^{\shortmid }\mathcal{E}$ or $\ \mathcal{E}=$ $\mathcal{E(M)}.$ In
formulas (\ref{dvbundle}), the group $\ ^{\shortmid }SO(4,1)$ is isomorphic
to $SO(4,1)$ but may have different parameterizations in variables on
typical co-fiber. The projection on base phase space $^{\shortmid }\pi $ can
be adapted to the N-connection splitting.

Let us consider orthonormalized frame transforms parameterized by some $%
8\times 8$ matrices $\ ^{\shortmid }\chi _{\ \alpha }^{\underline{\alpha }%
}(\ ^{\shortmid }u)\ $subjected to the condition that $\ ^{\shortmid }%
\mathbf{g}_{\alpha \beta }=\ ^{\shortmid }\chi _{\ \alpha }^{\underline{%
\alpha }}\ ^{\shortmid }\chi _{\ \beta }^{\underline{\beta }}\ ^{\shortmid
}\eta _{\underline{\alpha }\underline{\beta }},$ where the 8-d dubbing of
Minkovski metric can be written $\ ^{\shortmid }\eta _{\underline{\alpha }%
\underline{\beta }}=diag(1,1,1,-1,1,1,1,-1)$ in any point $\ ^{\shortmid
}u\in \ ^{\shortmid }\mathcal{M}.$ A canonical de Sitter nonliner gauge
gravitationl connection on $\ ^{\shortmid }\mathcal{E}$ is introduced as a
1-form 
\begin{equation}
\ ^{\shortmid }\widehat{\mathcal{A}}=\left[ 
\begin{array}{cc}
\ ^{\shortmid }\widehat{\mathcal{A}}_{\ \underline{\beta }}^{\underline{%
\alpha }} & l_{0}^{-1}\ ^{\shortmid }\chi ^{\underline{\alpha }} \\ 
l_{0}^{-1}\ ^{\shortmid }\chi _{\underline{\beta }} & 0%
\end{array}%
\right],  \label{dSgp}
\end{equation}%
where $l_{0}$ is a dimensional constant (it is necessary because the
dimensions of $\ ^{\shortmid }\widehat{\mathcal{A}}$- and $^{\shortmid }\chi 
$-fields different). In the matrix (\ref{dSgp}), we use $\ ^{\shortmid }%
\widehat{\mathcal{A}}_{\ \underline{\beta }}^{\underline{\alpha }}=\
^{\shortmid }\widehat{\mathcal{A}}_{\ \underline{\beta }\gamma }^{\underline{%
\alpha }}$ $\ ^{\shortmid }\mathbf{e}^{\gamma }=\ ^{\shortmid }\widehat{%
\mathcal{A}}_{\ \underline{\beta }\gamma _{s}}^{\underline{\alpha }}\
^{\shortmid }\mathbf{e}^{\gamma _{s}},$ see formulas (\ref{nadapb}), for $\
^{\shortmid }\widehat{\mathcal{A}}_{\ \underline{\beta }\gamma }^{\underline{%
\alpha }}=\ ^{\shortmid }\chi _{\ \alpha }^{\underline{\alpha }}\
^{\shortmid }\chi _{\underline{\beta }\ }^{\ \beta }\ ^{\shortmid }\widehat{%
\Gamma }_{\ \beta \gamma }^{\alpha }+\ ^{\shortmid }\chi _{\ \alpha }^{%
\underline{\alpha }}\ \ ^{\shortmid }\mathbf{e}_{\gamma }(\ ^{\shortmid
}\chi _{\underline{\beta }\ }^{\ \beta })$ determined by the N-, or
s-adapted coefficients of a canonical d-/s-connection $\ ^{\shortmid }%
\widehat{\mathbf{D}}$ (defined in Appendix \ref{appendixa}); and $\
^{\shortmid }\chi _{\ }^{\underline{\alpha }}=\ ^{\shortmid }\chi _{\ \alpha
}^{\underline{\alpha }}\ ^{\shortmid }\mathbf{e}^{\alpha }.$ Here we note
that we can consider similar constructions for an arbitrary linear
connection $\ ^{\shortmid }\Gamma _{\ \beta \gamma }^{\alpha },$ as in the
metric-affine geometry (in particular, for the Einstein-Cartan-Weyl spaces),
or a LC-connection $\ ^{\shortmid }\nabla ,$ but $\ _{s}^{\shortmid }%
\widehat{\mathbf{D}}$ has the priority to allow a general decoupling of the
phase space modified Einstein equations.

Using the Hodge operator $\divideontimes $ determined by $\ ^{\shortmid }%
\mathbf{g}_{\alpha \beta }$ and the absolute differential operator $\
^{\shortmid }d\ $ and skew product $\wedge $ on \ $\mathcal{\ ^{\shortmid }M}
$, we can define geometrically the curvature of (\ref{dSgp}), $\ ^{\shortmid
}\widehat{\mathcal{F}}=\ ^{\shortmid }d\ ^{\shortmid }\widehat{\mathcal{A}}%
+\ ^{\shortmid }\widehat{\mathcal{A}}$ $\wedge \ ^{\shortmid }\widehat{%
\mathcal{A}},$ and derive geometrically the commutative gauge gravitational
equations on $\ ^{\shortmid }\mathcal{E},$%
\begin{equation}
\ ^{\shortmid }d(\divideontimes \ ^{\shortmid }\widehat{\mathcal{F}})+\
^{\shortmid }\widehat{\mathcal{A}}\wedge (\divideontimes \ ^{\shortmid }%
\widehat{\mathcal{F}})-(\divideontimes \ ^{\shortmid }\widehat{\mathcal{F}}%
)\wedge \ ^{\shortmid }\widehat{\mathcal{A}}=-\lambda \ ^{\shortmid }%
\widehat{\mathcal{J}}.  \label{ymgreq}
\end{equation}%
The source in (\ref{ymgreq}) can be parameterized $\ ^{\shortmid }\widehat{%
\mathcal{J}}=\left[ 
\begin{array}{cc}
\ ^{\shortmid }\widehat{\mathcal{J}}_{\ \underline{\beta }}^{\underline{%
\alpha }} & -l_{0}t^{\underline{\alpha }} \\ 
-l_{0}t_{\underline{\beta }} & 0%
\end{array}%
\right] ,$ with $\ ^{\shortmid }\widehat{\mathcal{J}}_{\ \underline{\beta }%
}^{\underline{\alpha }}=\ ^{\shortmid }\widehat{\mathcal{J}}_{\ \underline{%
\beta }\gamma }^{\underline{\alpha }}$ $\ ^{\shortmid }\mathbf{e}^{\gamma }$
identified to zero for the model with LC-connection, or induced
nonholonomically for the canonical d-connection (we should consider it as a
spin density if we elaborate on phase space theories which are similar to
the Riemann-Cartan theory). The value $t^{\underline{\alpha }}=\ ^{\shortmid
}t_{\ \alpha }^{\underline{\alpha }}\ ^{\shortmid }\mathbf{e}^{\alpha }$ is
a phase space analog of the energy-momentum tensor for matter. The constant $%
\lambda $ can be related to the gravitational constant $l^{2}$ in 8-d (which
extends the 4-d one in GR) and other constants on phase space (from string
gravity etc.), when $l^{2}=2l_{0}^{2}\lambda ,\ \lambda _{1}=-3/l_{0}. $

Defining a s-adapted canonical gauge operator $\ ^{\shortmid }\widehat{%
\mathcal{D}}_{\alpha _{s}}:=$ $\ ^{\shortmid }\widehat{\mathbf{D}}_{\alpha
_{s}}+\ ^{\shortmid }\widehat{\mathcal{A}}_{\alpha _{s}}$ and with respect
to s-adapted frames (\ref{nadapb}) on \ $\ _{s}^{\shortmid }\mathcal{M},$ we
can writte the nonholonomic gauge gravitational field equations (\ref{nadapb}%
) in a form similar to the Yang-Mills, YM, equations,%
\begin{equation}
\ ^{\shortmid }\widehat{\mathcal{D}}_{\alpha _{s}}\ ^{\shortmid }\widehat{%
\mathcal{F}}^{\alpha _{s}\beta _{s}}=\ ^{\shortmid }\widehat{\mathbf{D}}%
_{\alpha _{s}}\ ^{\shortmid }\widehat{\mathcal{F}}^{\alpha _{s}\beta
_{s}}+[\ ^{\shortmid }\widehat{\mathcal{A}}_{\alpha _{s}},\ ^{\shortmid }%
\widehat{\mathcal{F}}^{\alpha _{s}\beta _{s}}]=-\lambda \ ^{\shortmid }%
\widehat{\mathcal{J}}^{\beta _{s}},  \label{ymgreq1}
\end{equation}%
for $[A,B]=AB-BA$ denoting the commutator on the Lie algebra of the chosen
gauge group$\ ^{\shortmid }\mathcal{G}r.$\footnote{%
Hereafter, we shall put a shell label $s,$ or use general indices of type $%
\alpha _{s},\beta _{s}$ $\ $in order to emphasize that such
geometric/physical \ s-objects and equations can be adapted to a necessary
type nonholonomic dyadic shell structure (being important for further
applications of the AFCDM) as we explain in Appendix \ref{appendixa}.} Such
equations have general decoupling and integration properties as in the case
of nonassociative modified Einstein equations from \cite{partner02} (we
shall discuss in section \ref{sec3}). The gauge gravitation fields on phase
space satisfy also the Bianchi identity (which is equivalent to the Jacoby
identity):%
\begin{equation}
\lbrack \ ^{\shortmid }\widehat{\mathcal{D}}_{\mu _{s}},[\ ^{\shortmid }%
\widehat{\mathcal{D}}_{\nu _{s}},\ ^{\shortmid }\widehat{\mathcal{D}}%
_{\alpha _{s}}]]+[\ ^{\shortmid }\widehat{\mathcal{D}}_{\alpha _{s}},[\
^{\shortmid }\widehat{\mathcal{D}}_{\mu _{s}},\ ^{\shortmid }\widehat{%
\mathcal{D}}_{\nu _{s}}]]+[\ ^{\shortmid }\widehat{\mathcal{D}}_{\nu
_{s}},[\ ^{\shortmid }\widehat{\mathcal{D}}_{\alpha _{s}},\ ^{\shortmid }%
\widehat{\mathcal{D}}_{\mu _{s}}]]=0,  \label{jakoby}
\end{equation}%
considered for matrix operators with values in Lie algebra.

\subsection{Nonassociative gauge gravity for the double de Sitter group}

Applying on a phase space and gauge geometric objects the nonassociative
star product operator $\star _{s}$ (see formula (\ref{starpn}), determined
by R-flux deformations adapted to the dyadic shell structure $s=1,2,3,4$ on $%
\ _{s}^{\shortmid }\mathcal{M}\mathbf{\rightarrow }\ _{s}^{\shortmid }%
\mathcal{M}^{\star }),$ we can define the R--deformed nonassociative
geometric and physical objects considered for the above gauge gravity model.
On a base phase space, $\star _{s}:\ \mathbf{g}_{s}\mathbf{\rightarrow g}%
_{s}^{\star }= (\breve{g}_{s}^{\star }, \check{g}_{s}^{\star });\ \
_{s}^{\shortmid }\widehat{\mathbf{D}}= \ _{s}^{\shortmid }\nabla + \
_{s}^{\shortmid }\widehat{\mathbf{Z}} \rightarrow \ _{s}^{\shortmid }%
\widehat{\mathbf{D}}^{\star }=\ _{s}^{\shortmid }{\nabla }^{\star }+\
_{s}^{\shortmid }\widehat{\mathbf{Z}}^{\star}; \divideontimes \rightarrow 
\breve{\divideontimes},$ when the Hodge operator \ $\breve{\divideontimes}$
on $\ _{s}^{\shortmid }\mathcal{M}^{^{\star }}$ can be defined using the
symmetric part $\breve{g}_{s}^{\star }$ of the star product deformed
s-metric. The nonholonomic dyadic approach allows us to compute the
antisymmetric part of the star deformed metric, $\mathbf{\check{g}}%
_{s}^{\star },$ and induced by R-flux term (this is possible if we apply the
Convention 2 from \cite{partner02}\ for computing such $\star _{s}$%
-deformations in explicit s-adapted form).

The (co) vector bundles on $\ _{s}^{\shortmid }\mathcal{M}$, subjected to $%
\star _{s}$-deformations transform into respective nonassociative spaces,
for instance, $\ \ _{s}^{\shortmid }\mathcal{E(}\ _{s}^{\shortmid }\mathcal{M%
})\rightarrow \ \ _{s}^{\shortmid }\mathcal{E}^{\star }(\ _{s}^{\shortmid }%
\mathcal{M}^{\star })= (\ _{s}^{\shortmid }\mathcal{E}^{\star },\
^{\shortmid }\mathcal{G}r,\ ^{\shortmid }\pi , \ _{s}^{\shortmid }\mathcal{M}%
^{\star }),$ when the double structure group $\ ^{\shortmid }\mathcal{G}r$
is preserved as in (\ref{dvbundle}) but $\star _{s}$-deformations are
considered for the base phase space and s-adapted components of the
geometric objects.\footnote{%
We can elaborate on more general classes of nonassociative gauge models
when, for instance, $\ ^{\shortmid }\mathcal{G}r$ transforms into some
quantum groups (additionally to $\star _{s}$-deformations). In general, such
theories involve various types of algebraic assumptions and request new
physical motivations comparing to the "nonassociative string theory R-flux
deformation philosophy; and it is not clear how to prove any general
decoupling and integration properties for such quantum gauge gravitational
models. We restrict our research program outlined in \cite%
{partner02,partner04} only to nonassociative gravitational and matter field
theories when the gauge groups are not subjected to quantum deformation. In
such cases, we can apply the AFCDM and generate exact and parametric
solutions of physically important systems of nonlinear PDEs.} In a similar
abstract (and nonholonomic s-adapted) geometric form, we define and compute
s-adapted components for: $\star _{s}:\ ^{\shortmid }\widehat{\mathcal{A}}
\rightarrow \ _{s}^{\shortmid }\widehat{\mathcal{A}} ^{\star },$ with $\
^{\shortmid }\mathbf{g}_{\alpha \beta }$ identified to $\breve{g}_{s}^{\star
}$ is we consider zero powers of parameters $\hbar $ and $\kappa $ from (\ref%
{starpn}), which allows to use the same $\ ^{\shortmid }\chi ^{\underline{%
\alpha }}$ and $\ ^{\shortmid }\mathbf{e}^{\gamma _{s}}$ from/for (\ref{dSgp}%
), but $\star _{s}:\ \ ^{\shortmid }\widehat{\mathcal{A}}_{\ \underline{%
\beta }\gamma _{s}}^{\underline{\alpha }}\rightarrow :\ \ ^{\shortmid }%
\widehat{\mathcal{A}}_{\ \underline{\beta }\gamma _{s}}^{\star \underline{%
\alpha }}.$ Then, we can compute nonassociative star product deformations of
type $\star _{s}:\ ^{\shortmid }\widehat{\mathcal{F}}= \ ^{\shortmid }d\
^{\shortmid }\widehat{\mathcal{A}}+\ ^{\shortmid }\widehat{\mathcal{A}}
\wedge \ ^{\shortmid}\widehat{\mathcal{A}}\rightarrow \ _{s}^{\shortmid }%
\widehat{\mathcal{F}}^{\star }= \ _{s}^{\shortmid }d\ \ _{s}^{\shortmid }%
\widehat{\mathcal{A}}^{\star }+ \ _{s}^{\shortmid }\widehat{\mathcal{A}}$ $%
^{\star }\wedge ^{s\star }\ \ _{s}^{\shortmid }\widehat{\mathcal{A}}^{\star
},$ with a respective s-adapted anti-symmetric operator $\wedge \rightarrow
\wedge ^{s\star };$ and, for the generalized source of the YM equations, $%
\star _{s}:\ ^{\shortmid }\widehat{\mathcal{J}}^{\beta _{s}}\rightarrow \
^{\shortmid }\widehat{\mathcal{J}}^{\star \beta _{s}}.$ On the coefficient
definition of the star product deformations of commutator, see \cite%
{partner02,blumenhagen16,aschieri17}, which allows us to compute $%
[A,B]\rightarrow \lbrack A^{\star },B^{\star }]^{\star }$ etc.

In abstract star product geometric form, we can follow the above stated
rules and introduce and compute $\ _{s}^{\shortmid }\widehat{\mathcal{D}}=\
_{s}^{\shortmid }\widehat{\mathbf{D}}+\ _{s}^{\shortmid }\widehat{\mathcal{A}%
}\rightarrow \ _{s}^{\shortmid }\widehat{\mathcal{D}}^{\star }=\
_{s}^{\shortmid }\widehat{\mathbf{D}}^{^{\star }}+ \ _{s}^{\shortmid }%
\widehat{\mathcal{A}}^{^{\star }},$ and $\ _{s}^{\shortmid }\widehat{%
\mathcal{J}}\rightarrow \ _{s}^{\shortmid }\widehat{\mathcal{J}}^{\star }.$
This allows to $\star _{s}$-deform the YM type equations for the de Sitter
phase space gravity (\ref{ymgreq}) and/or (\ref{ymgreq1}), $\ ^{\shortmid }d(%
\breve{\divideontimes}\ ^{\shortmid }\widehat{\mathcal{F}}^{^{\star }})+\
^{\shortmid }\widehat{\mathcal{A}}^{^{\star }}\wedge (\divideontimes \
^{\shortmid }\widehat{\mathcal{F}}^{^{\star }})-(\divideontimes \
^{\shortmid }\widehat{\mathcal{F}}^{^{\star }})\wedge \ ^{\shortmid }%
\widehat{\mathcal{A}}^{^{\star }}=-\lambda \ ^{\shortmid }\widehat{\mathcal{J%
}}^{^{\star }}$, which can be written in nonholonomic s-adapted form, 
\begin{equation}
\ ^{\shortmid }\widehat{\mathcal{D}}_{\alpha _{s}}^{^{\star }}\ ^{\shortmid }%
\widehat{\mathcal{F}}^{^{\star }\alpha _{s}\beta _{s}}=\ ^{\shortmid }%
\widehat{\mathbf{D}}_{\alpha _{s}}^{^{\star }}\ ^{\shortmid }\widehat{%
\mathcal{F}}^{^{\star }\alpha _{s}\beta _{s}}+[\ ^{\shortmid }\widehat{%
\mathcal{A}}_{\alpha _{s}}^{^{\star }},\ ^{\shortmid }\widehat{\mathcal{F}}%
^{^{\star }\alpha _{s}\beta _{s}}]^{\star }=-\lambda \ ^{\shortmid }\widehat{%
\mathcal{J}}^{^{\star }\beta _{s}}.  \label{nonassocymgreq1}
\end{equation}%
Here we note that $\star _{s}$-deforms of the Bianchi identities (\ref%
{jakoby}) result, in general, in a nonzero "Jakobiator" (with nontrivial
right part), which is typical for nonassociative theories (we omit explicit
formulas). Such values can be computed as induced ones by choosing
respective s-adapted configurations with $\hbar $ and $\kappa $ parametric
decompositions.

Finally, we conclude that if an effective or matter field source $\
^{\shortmid }\widehat{\mathcal{J}}^{^{\star }\beta _{s}}$ in (\ref%
{nonassocymgreq1}) is correspondingly parameterized and physically
motivated, such nonassociative YM equations present alternatives and
generalizations of the nonassociative star product deformed Einstein
equations considered in vacuum form in \cite{blumenhagen16,aschieri17} and,
in s-adapted form, with extensions to certain classes of nontrivial sources, 
\cite{partner02,partner04}.

\subsection{Projecting nonassociative gravitational YM eqs into
nonassociative Einstein eqs}

We can consider instead of the de Sitter structure group $SO(4,1)$ the
affine structure group $Af(4,1)$, when the gauge potential for our theory on
phase space $\ _{s}^{\shortmid }\mathcal{M}^{\star }$ takes values into the
double Poincar\'{e}-Lie algebra (\ref{dSgp}),%
\begin{equation}
\ _{s}^{\shortmid }\widehat{\mathcal{A}}^{\star }\rightarrow \
_{s}^{\shortmid }\widehat{\mathcal{A}}_{[P]}^{\star }=\left[ 
\begin{array}{cc}
\ _{s}^{\shortmid }\widehat{\mathcal{A}}_{\ \underline{\beta }}^{\star 
\underline{\alpha }} & l_{0}^{-1}\ _{s}^{\shortmid }\chi ^{\underline{\alpha 
}} \\ 
\chi _{0}^{\underline{\alpha }} & 0%
\end{array}%
\right] ; \mbox{ constants } \chi _{0}^{\underline{\alpha }}%
\mbox{ can be
chosen to be }0 \mbox{ at the end of certain computations}.  \label{affinpot}
\end{equation}%
Then, we can introduce a source $\ _{s}^{\shortmid }\widehat{\mathcal{J}}%
^{\star }=\left[ 
\begin{array}{cc}
\ _{s}^{\shortmid }\widehat{\mathcal{J}}_{\ \underline{\beta }}^{\star 
\underline{\alpha }}=0 & -l_{0}\ _{s}^{\shortmid }t^{\star \underline{\alpha 
}} \\ 
\chi _{0}^{\underline{\alpha }} & 0%
\end{array}%
\right] ,$ $\ _{s}^{\shortmid }t^{\star \underline{\alpha }}=\chi ^{%
\underline{\alpha }\beta _{s}}\ \ ^{\shortmid }\Im _{\alpha _{s}\beta
_{s}}^{\star }\ ^{\shortmid }\mathbf{e}^{\alpha },$ where $\ ^{\shortmid
}\Im _{\alpha _{s}\beta _{s}}^{\star }$ is the star product deformation of
the effective energy-momentum tensor written in s-adapted form $\
^{\shortmid }\Im _{\alpha _{s}\beta _{s}}$ on $\ _{s}^{\shortmid }\mathcal{M}%
^{\star }.$ Such nonholonomic nonassociative and commutative sources are
considered in nonassociative gravity and nonassociative geometric flow
theories. For the conditions, the nonassociative YM equations (\ref%
{nonassocymgreq1}) transform into nonassociative s-adapted canonical
gravitational equations from \cite{partner02,partner04},%
\begin{equation}
\ ^{\shortmid }\widehat{\mathbf{R}}ic_{\alpha _{s}\beta _{s}}^{\star }=\
^{\shortmid }\Im _{\alpha _{s}\beta _{s}}^{\star }.  \label{nonassocaneinst}
\end{equation}%
In the vacuum case with $^{\shortmid }\Im _{\alpha _{s}\beta _{s}}^{\star
}=0 $ and $\ _{s}^{\shortmid }\widehat{\mathbf{D}}^{\star }\rightarrow \
_{s}^{\shortmid }{\nabla }^{\star },$ we obtain the vacuum gravitational
equations for nonassociative and noncommutative gravity studied in \cite%
{blumenhagen16,aschieri17}. More than that, if we construct the affine
potential (\ref{affinpot}) for the standard LC-connection ${\nabla }$ on a
pseudo-Riemannian spacetime base, both the equations (\ref{nonassocymgreq1})
and (\ref{nonassocaneinst}) transform into the standard Einstein equations
in GR written in abstract form as in \cite{misner}. Above equations can be
proven in s-adapted form a tedious calculus considered in \cite%
{svnc00,vd00,sv00} for nonholonomic commutative phase spaces and in
noncommutative gauge gravity with Seiberg-Witten product.

Let us discuss the (non) variational properties of the nonassociative YM
equations (\ref{nonassocymgreq1}) derived following "pure" geometric
methods. It is a well-known fact, that it is not possible to elaborate a
unique and self-consistent variational formalism involving general twisted
products but we can always follow abstract/ symbolic geometric proofs. The
main reason is that we can introduce an infinite number of nonassociative
and noncommutative differential and integral calculi in abstract form but
have to elaborate on explicit differentials and integrals when a star
deformation structure is stated in a unique form. Nevertheless, both classes
of nonassociative systems of PDEs (\ref{nonassocymgreq1}) or (\ref%
{nonassocaneinst}) are mathematically well-defined, with self-consistent and
physically important solutions if certain nonholonomic assumptions are
stated:\ If we begin with a variational theory on the phase space (related,
for instance, to variational Einstein equations on the base spacetime, see 
\cite{svnc00,sv00}), then deform it using the nonassociative star product
(for applications in MGTs, we can consider linear approximations on $\hbar $
and $\kappa $ parameters. Such classical nonassociative models with
effective sources can be written as effective theories with a prescribed
variational procedure involving $\star _{s}$-deformations. It is possible to
construct nonassociative and noncommutative versions of R-flux deformed
black hole/ellipsoid, toroid, wormhole and cosmological solutions as in \cite%
{svnc00,vd00,sv00,sv15a,partner02,partner04}.

Another issue related to non-variational theories is that the affine and
Poincare structure groups are not semi-simple and such theories are not
variational in the respective total spaces. This can be avoided if we work
with gauge potentials of type (\ref{affinpot}) for nontrivial constants $%
\chi _{0}^{\underline{\alpha }}.$ In such cases, the commutative part of
such theories is positively variational \ and we can consider $\chi _{0}^{%
\underline{\alpha }}\rightarrow 0$ after some physically important solutions
are zero. In explicit form, we can also consider nonlinear enveloping
algebras and nonlinear gauge group realizations as in \cite{jurco,svnc00}.
The extensions of type $Af(4,1)\rightarrow $ $SO(4,1)$ generalize the
theories both in commutative and nonassociative/ noncommutative forms when
the YM gravitational equations (\ref{nonassocymgreq1}) became more general
than (\ref{nonassocaneinst}) and allows to introduce additional sources $\
^{\shortmid }\widehat{\mathcal{J}}_{\ \underline{\beta }}^{\star \underline{%
\alpha }}$ for more general classes of nontrivial torsion fields (not only
those nonholonomically induced).

\section{Off-diagonal cosmological solutions in nonassociative gravity}

\label{sec3} We can construct various classes of exact and parametric
solutions of the nonassociative gauge YM equations (\ref{nonassocymgreq1})
if we are able to find solutions of the nonassociative Einstein equations
written in canonical dyadic variables (\ref{nonassocaneinst}). Such systems
of nonlinear PDEs possess a general decoupling property if the source of
effective and matter fields can be parameterized (using respective
nonholonomic frame transforms and connection deformations) as $\tau $%
--families of s-adapted coefficients $\ ^{\shortmid }\Im _{\ \ \beta
_{s}}^{\star \alpha _{s}}(\tau )=[~_{1}^{\shortmid }\mathcal{K}(\tau )\delta
_{i_{1}}^{j_{1}},~_{2}^{\shortmid }\mathcal{K}(\tau )\delta
_{b_{2}}^{a_{2}},~_{3}^{\shortmid }\mathcal{K}(\tau )\delta
_{a_{3}}^{b_{3}},~_{4}^{\shortmid }\mathcal{K}(\tau )\delta
_{a_{4}}^{b_{4}}] $ (\ref{cannonsymparamc2a}), see Appendix \ref{appendixa}
and details on the AFCDM in \cite{partner02,partner04}.\footnote{%
We can consider $0\leq \tau \leq \tau _{0}$ as a temperature like real
parameter, which is important for elaborating thermodynamic cosmological
models in next section. Here we also note that the label $\ "^{\shortmid }"$
refers to real phase space coordinates, when $\ "^{\shortparallel }"$ is
used for complexified momenta coordinates which for decompositions on the
string constant $\kappa $ result in real additional terms encoding
nonassociative R-flux terms.}

To generate off-diagonal cosmological solutions we can take any
quasi-stationary metric constructed in \cite{partner04} and subject it to
so-called time-space duality transforms when, for instance, $%
x^{3}\rightarrow x^{4}=t$ and, in our work, indices change mutually as $%
3\rightarrow 4$ and $4\rightarrow 3.$ For applications in modern cosmology,
it is important to study $\tau $-families of solutions describing
nonassociative deformations of a FLRW metric, $\mathbf{\mathring{g}}=[%
\mathring{g}_{\alpha },\ \mathring{N}_{i}^{a}],$ into solutions $\mathbf{g}%
_{s}^{\star }(\tau _{0})=[\breve{g}_{s}^{\star }(\tau ,x^{i},t,p_{a_{s}}),%
\check{g}_{s}^{\star }(\tau ,x^{i},t,p_{a_{s}})]$ of nonassociative
cosmological Ricci soliton equations 
\begin{equation}
\ ^{\shortmid }\widehat{\mathbf{R}}ic_{\alpha _{s}\beta _{s}}^{\star }(\tau
_{0},x^{i},t,p_{a_{s}})=\ _{s}^{\shortmid }\Lambda (\tau _{0})\ \mathbf{%
\breve{g}}_{\alpha _{s}\beta _{s}}^{\star }(\tau _{0},x^{i},t,p_{a_{s}}).
\label{nariccisol}
\end{equation}%
Such systems of nonlinear PDEs can be derived for self-similar
configurations for any fixed $\tau _{0}$ in the nonholonomic geometric flow
equations considered in various (non) associative / commutative theories 
\cite{perelman1,svnonh08,partner04,kehagias19}.\footnote{%
A prime off-diagonal metric $\mathbf{\mathring{g}}$ is extended trivially
from 4-d and 8-d phase spaces, i.e. from 2 shells to 4 shells, a $\mathbf{%
\mathring{g}}_{\alpha _{2}\beta _{2}}$ is written using general spacetime
curved coordinates $u^{\alpha _{2}}=u^{\alpha _{2}}(x,y,z,t)$ ( for $%
u^{1}=x,u^{2}=y,u^{3}=t,u^{4}=t$ with $\alpha _{2}=i_{1},a_{2},$ when $%
i_{1}=1,2$ and $a_{2}=3,4)$ and considered as a conformal transform, $%
\mathring{g}_{\alpha \beta }\simeq \mathring{a}^{-2}(t)\ ^{RW}\mathbf{g}%
_{\alpha \beta },$ of the Friedmann-Lema\^{\i}tre-Robertson-Walker, FLRW,
diagonal metric, $d\mathring{s}^{2}=\ ^{RW}\mathbf{g}_{\alpha \beta }(u)%
\mathbf{e}^{\alpha }\mathbf{e}^{\beta }\simeq \mathring{a}%
^{2}(t)(dx^{2}+dy^{2}+dz^{2})-dt^{2}.$ In these formulas, $t$ is the cosmic
time, $\mathring{a}(t)$ is the scale factor; and $x^{\grave{\imath}}=(x,y,z)$
are the Cartesian coordinates.} The Ricci soliton cosmological solutions of (%
\ref{nariccisol}) can be related to the solutions of (\ref{nonassocaneinst})
and (\ref{nonassocymgreq1}) via some nonlinear symmetries 
\begin{eqnarray}
(\ _{s}\Psi ,\ ~_{s}^{\shortmid }\mathcal{K}) &\leftrightarrow &(\ _{s}%
\widehat{\mathbf{g}}=\breve{g}_{s}^{\star }(\tau ,x^{i},t,p_{a_{s}}),\
~_{s}^{\shortmid }\mathcal{K})\leftrightarrow (~_{s}^{\shortmid }\eta \ \ \
\ _{s}^{\shortmid }\mathring{g}_{\alpha _{s}}\sim \ _{s}^{\shortmid }\zeta
(1+\kappa \ _{s}^{\shortmid }\chi _{\alpha _{s}})\ \ _{s}^{\shortmid }%
\mathring{g}_{\alpha _{s}},\ ~_{s}^{\shortmid }\mathcal{K})\leftrightarrow
\label{nonlinsym} \\
(\ _{s}\Phi ,\ _{s}\Lambda _{0}) &=&\ _{s}^{\shortmid }\Lambda (\tau
_{0}))\leftrightarrow (\ _{s}\widehat{\mathbf{g}},\ \ _{s}\Lambda
_{0})\leftrightarrow (~_{s}^{\shortmid }\eta \ \ \ \ _{s}^{\shortmid }%
\mathring{g}_{\alpha _{s}}\sim \ _{s}^{\shortmid }\zeta (1+\kappa \
_{s}^{\shortmid }\chi _{\alpha _{s}})\ \ _{s}^{\shortmid }\mathring{g}%
_{\alpha _{s}},\ \ _{s}^{\shortmid }\Lambda _{0}).  \notag
\end{eqnarray}%
Explicit formulas and details on such nonlinear symmetries and respective
formulas for Killing symmetries on $\partial _{4}$ are provided below, at
the end of section 2 and in appendix A.2 to \cite{partner02,partner04}. We
note that for the purposes of this work one has to change the Killing
symmetries on $\partial _{3}$ into those on $\partial _{4}=\partial t$. In (%
\ref{nonlinsym}), $\ _{s}^{\shortmid }\mathring{g}_{\alpha _{s}}$ are
considered a prime s-metrics (for our purposes, we can chose a FLRW one);
there are used generation functions $\ _{s}^{\shortmid }\Psi (\tau
,x^{i},t,p_{a_{s}}),$ or$\ \ _{s}^{\shortmid }\Phi (\tau
,x^{i},t,p_{a_{s}}), $ and respective graviational polarization functions $%
~_{s}^{\shortmid }\eta (\tau ,x^{i},t,p_{a_{s}}),$ or $\ _{s}^{\shortmid
}\zeta (\tau ,x^{i},t,p_{a_{s}})$ and $\ _{s}^{\shortmid }\chi _{\alpha
_{s}}(\tau ,x^{i},t,p_{a_{s}});$ when $~_{s}^{\shortmid }\mathcal{K}(\tau
,x^{i},t,p_{a_{s}})$ are considered as generating sources and $\
_{s}^{\shortmid }\Lambda (\tau )$ are effective $\tau $-running cosmological
constants. Such generating data together with integration functions and
constants for (\ref{nariccisol}) can be chosen in certain forms which, for
instance, may describe new obsevable JWST cosmological data, and reproduce
certain propertied of the DM and DE physics as explained in \cite%
{foroconi23,boylan23,biagetti23}.

\subsection{Cosmological solitonic hierarchies in nonassociative gravity}

\label{ssec31}Applying the AFCDM, we can construct such $\tau $-families of
generic off-diagonal cosmological solutions defining nonassociative phase
space deformations of FLRW metrics: 
\begin{eqnarray}
d\widehat{s}^{2}(\tau ) &=&e^{\psi (\tau )}[(dx^{1})^{2}+(dx^{2})^{2}]+\eta
_{3}(\tau )\mathring{g}_{3}[dy^{3}+n_{k_{1}}(\tau )dx^{k_{1}}]+\eta
_{4}(\tau )\mathring{g}_{4}[dt+w_{k_{1}}(\tau )dx^{k_{1}}]  \label{cosmoffds}
\\
&&+\ ^{\shortmid }\eta ^{5}(\tau )\ ^{\shortmid }\mathring{g}^{5}[dp_{5}+\
^{\shortmid }n_{k_{2}}(\tau )d\ ^{\shortmid }x^{k_{2}}]+\ ^{\shortmid }\eta
^{6}(\tau )\ ^{\shortmid }\mathring{g}^{6}[dp_{6}+\ ^{\shortmid
}w_{k_{2}}(\tau )d\ ^{\shortmid }x^{k_{2}}]  \notag \\
&&+\ ^{\shortmid }\eta ^{7}(\tau )\ ^{\shortmid }\mathring{g}^{7}[dp_{7}+\
^{\shortmid }n_{k_{3}}(\tau )d\ ^{\shortmid }x^{k_{3}}]+\ ^{\shortmid }\eta
^{8}(\tau )\ ^{\shortmid }\mathring{g}^{8}[dE+\ ^{\shortmid }w_{k_{3}}(\tau
)d\ ^{\shortmid }x^{k_{3}}],  \notag
\end{eqnarray}%
where $g_{1}=e^{\psi (\tau )}\mathring{g}_{1},g_{2}=e^{\psi (\tau )}%
\mathring{g}_{2};$ for $\mathring{g}_{1}=\mathring{g}_{2}=\mathring{g}_{3}=1,%
\mathring{g}_{4}=-\mathring{a}^{-2}(t),\ ^{\shortmid }\mathring{g}^{5}=\
^{\shortmid }\mathring{g}^{6}=\ ^{\shortmid }\mathring{g}^{7}=1,\
^{\shortmid }\mathring{g}^{8}=-1;$ and indices and local phase space
coordinates labelled as $k=k_{1}=1,2;k_{2}=1,2,3,4;k_{3}=1,2,...6;\
^{\shortmid }u^{\alpha
_{s}}=(x^{k_{1}},y^{3}=x^{3},y^{4}=x^{4}=t,p_{5},p_{6},p_{7},p_{8}=E),$ see
conventions (\ref{coordconv}). To generate solutions of (\ref%
{nonassocaneinst}) the coefficients in (\ref{cosmoffds}) can be defined in
such functional forms:

For the s-adapted metric coefficients,%
\begin{eqnarray}
\psi (\tau ) &=&\psi (\tau ,x^{k})%
\mbox{ are solutions of 2-d Poisson
equations }\partial _{1}^{2}\psi +\partial _{2}^{2}\psi =~_{1}^{\shortmid }%
\mathcal{K}(\tau ,x^{k});  \label{smetrc} \\
\eta _{3}(\tau ) &=&\eta _{3}[\tau ,\ ^{\widehat{a}}\wp (\tau ,x^{i},t)]%
\mbox{ is a family of generating functions / functionals };  \notag \\
\eta _{4}(\tau ) &=&\eta _{4}[\tau ,\ ^{\widehat{a}}\wp (\tau ,x^{i},t),\ ^{%
\widehat{a}}\xi (\tau ,x^{i},t)]=-\frac{[\partial _{t}(\ \eta _{3}\ 
\mathring{g}_{3})]^{2}}{|\int dt\ ~_{2}\mathcal{K}(\tau ,\ ^{\widehat{a}}\wp
,\ ^{\widehat{a}}\xi )\partial _{t}(\ \eta _{3}\ \mathring{g}_{3})|\ \eta
_{3}\mathring{g}_{3}};\   \notag
\end{eqnarray}%
\begin{eqnarray*}
\ ^{\shortmid }\eta ^{5}(\tau ) &=&\ ^{\shortmid }\eta ^{5}[\tau ,\ \ ^{%
\widehat{a}}\wp (\tau ,x^{i},t)]%
\mbox{ is a family of generating functions /
functionals }; \\
\ ^{\shortmid }\eta ^{6}(\tau ) &=&\ ^{\shortmid }\eta ^{6}[\tau ,\ ^{%
\widehat{a}}\wp (\tau ,x^{i},t),\ ^{\widehat{a}}\xi (\tau ,x^{i},t)]=-\frac{%
[\ ^{\shortmid }\partial ^{6}(\ ^{\shortmid }\eta ^{5}\ ^{\shortmid }%
\mathring{g}^{5})]^{2}}{|\int dp_{6}\ ~_{3}^{\shortmid }\mathcal{K}(\tau ,\
^{\widehat{a}}\wp ,\ ^{\widehat{a}}\xi )\ ^{\shortmid }\partial ^{6}(\
^{\shortmid }\eta ^{5}\ ^{\shortmid }\mathring{g}^{5})|\ ^{\shortmid }\eta
^{5}\ ^{\shortmid }\mathring{g}^{5}};
\end{eqnarray*}%
\begin{eqnarray*}
\ ^{\shortmid }\eta ^{7}(\tau ) &=&\ ^{\shortmid }\eta ^{7}[\tau ,\ \ ^{%
\widehat{a}}\wp (\tau ,x^{i},t)]%
\mbox{ is a family of generating functions /
functionals }; \\
\ ^{\shortmid }\eta ^{8}(\tau ) &=&\ ^{\shortmid }\eta ^{8}[\tau ,\ ^{%
\widehat{a}}\wp (\tau ,x^{i},t),\ ^{\widehat{a}}\xi (\tau ,x^{i},t)]=-\frac{%
[\ ^{\shortmid }\partial ^{8}(\ ^{\shortmid }\eta ^{7}\ ^{\shortmid }%
\mathring{g}^{7})]^{2}}{|\int dE\ ~_{3}^{\shortmid }\mathcal{K}(\tau ,\ ^{%
\widehat{a}}\wp ,\ ^{\widehat{a}}\xi )\ \ ^{\shortmid }\partial ^{8}(\
^{\shortmid }\eta ^{7}\ ^{\shortmid }\mathring{g}^{7})|\ ^{\shortmid }\eta
^{7}\ ^{\shortmid }\mathring{g}^{7}}.
\end{eqnarray*}%
For the N-connection coefficients, see (\ref{scon}) and (\ref{nadapb}), we
have such $\tau $-parametric and functional dependencies: 
\begin{eqnarray}
n_{k_{1}}(\tau ) &=&\ _{1}n_{k_{1}}(\tau )+\ _{2}n_{k_{1}}(\tau )\int dt%
\frac{[\partial _{t}(\eta _{3}\mathring{g}_{3})]^{2}}{|\int dt\ ~_{2}%
\mathcal{K}(\tau ,\ ^{\widehat{a}}\wp ,\ ^{\widehat{a}}\xi )\partial
_{t}(\eta _{3}\mathring{g}_{3})|\ (\eta _{3}\mathring{g}_{3})^{5/2}},  \notag
\\
w_{k_{1}}(\tau ) &=&\frac{\partial _{k_{1}}[\int dt\ ~_{2}\mathcal{K}(\tau
,\ ^{\widehat{a}}\wp ,\ ^{\widehat{a}}\xi )\ \ \partial _{t}(\eta _{3}%
\mathring{g}_{3})]}{\ ~_{2}\mathcal{K}(\tau ,\ ^{\widehat{a}}\wp ,\ ^{%
\widehat{a}}\xi )\partial _{t}(\eta _{3}\mathring{g}_{3})};  \label{nconc}
\end{eqnarray}%
\begin{eqnarray*}
\ ^{\shortmid }n_{k_{2}}(\tau ) &=&\ _{1}^{\shortmid }n_{k_{2}}(\tau )+\
_{2}^{\shortmid }n_{k_{2}}(\tau )\int dp_{6}\frac{[\ ^{\shortmid }\partial
^{6}(\ ^{\shortmid }\eta ^{5}\ ^{\shortmid }\mathring{g}^{5})]^{2}}{|\int
dp_{6}\ \ ~_{3}^{\shortmid }\mathcal{K}(\tau ,\ ^{\widehat{a}}\wp ,\ ^{%
\widehat{a}}\xi )\ ^{\shortmid }\partial ^{6}(\ ^{\shortmid }\eta ^{5}\
^{\shortmid }\mathring{g}^{5})|\ (\ ^{\shortmid }\eta ^{5}\ ^{\shortmid }%
\mathring{g}^{5})^{5/2}}, \\
\ ^{\shortmid }w_{k_{2}}(\tau ) &=&\frac{\partial _{k_{2}}[\int dp_{6}\
~~_{3}^{\shortmid }\mathcal{K}(\tau ,\ ^{\widehat{a}}\wp ,\ ^{\widehat{a}%
}\xi )\ ^{\shortmid }\partial ^{6}(\ ^{\shortmid }\eta ^{5}\ ^{\shortmid }%
\mathring{g}^{5})]}{\ ~_{3}^{\shortmid }\mathcal{K}(\tau ,\ ^{\widehat{a}%
}\wp ,\ ^{\widehat{a}}\xi )\ ^{\shortmid }\partial ^{6}(\ ^{\shortmid }\eta
^{5}\ ^{\shortmid }\mathring{g}^{5})};
\end{eqnarray*}%
\begin{eqnarray*}
\ ^{\shortmid }n_{k_{3}}(\tau ) &=&\ _{1}^{\shortmid }n_{k_{2}}(\tau )+\
_{2}^{\shortmid }n_{k_{2}}(\tau )\int dE\frac{[\ ^{\shortmid }\partial
^{8}(\ ^{\shortmid }\eta ^{7}\ ^{\shortmid }\mathring{g}^{7})]^{2}}{|\int
dE\ \ ~_{4}^{\shortmid }\mathcal{K}(\tau ,\ ^{\widehat{a}}\wp ,\ ^{\widehat{a%
}}\xi )\ ^{\shortmid }\partial ^{8}(\ ^{\shortmid }\eta ^{7}\ ^{\shortmid }%
\mathring{g}^{7})|\ (\ ^{\shortmid }\eta ^{7}\ ^{\shortmid }\mathring{g}%
^{7})^{5/2}}, \\
\ ^{\shortmid }w_{k_{3}}(\tau ) &=&\frac{\partial _{k_{3}}[\int dE\
~~_{4}^{\shortmid }\mathcal{K}(\tau ,\ ^{\widehat{a}}\wp ,\ ^{\widehat{a}%
}\xi )\ ^{\shortmid }\partial ^{8}(\ ^{\shortmid }\eta ^{7}\ ^{\shortmid }%
\mathring{g}^{7})]}{\ ~_{4}^{\shortmid }\mathcal{K}(\tau ,\ ^{\widehat{a}%
}\wp ,\ ^{\widehat{a}}\xi )\ ^{\shortmid }\partial ^{8}(\ ^{\shortmid }\eta
^{7}\ ^{\shortmid }\mathring{g}^{7})}.
\end{eqnarray*}%
In these formulas, there are considered $\tau $-families of integration
functions$\ _{1}n_{k_{1}}(\tau )=\ _{1}n_{k_{1}}(\tau ,x^{i_{1}}),\ \
_{2}n_{k_{1}}(\tau )=\ _{2}n_{k_{1}}(\tau ,x^{i_{1}});$ $_{1}n_{k_{2}}(\tau
)=\ _{1}n_{k_{2}}(\tau ,x^{i_{1}},y^{a_{2}}),\ \ _{2}n_{k_{2}}(\tau )=\
_{2}n_{k_{2}}(\tau ,x^{i_{1}},y^{a_{2}});$ and $_{1}n_{k_{3}}(\tau )=\
_{1}n_{k_{3}}(\tau ,x^{i_{1}},y^{a_{2}},p_{a_{3}}),$ \newline
$\ \ _{2}n_{k_{3}}(\tau )=\ _{2}n_{k_{3}}(\tau
,x^{i_{1}},y^{a_{2}},p_{a_{3}}).$ Such values can be chosen to describe
certain observational cosmological data.

Any $^{\shortmid }\mathbf{g}(\tau )=\ ^{\shortmid }g[\ ^{\widehat{a}}\wp ,\
^{\widehat{a}}\xi ]$ (\ref{cosmoffds}) defines $\tau $-families of
cosmological solitonic hierarchies if the s-adapted coeficients of s-metric (%
\ref{smetrc}) \ and N-connection (\ref{nconc}) are generatied as functionals
on solitonic waves (for instance, of type (\ref{solitdistr})) as explained
in Apppendix \ref{appendixb}. Such generic off-diagonal and locally
anisotropic cosmological metrics define exact and parametric solitonic
solutions of nonassociative gauge gravity YM \ (\ref{nonassocymgreq1})
and/or nonassociative Einstein (\ref{nonassocaneinst}) equations if the
generating sources $\ ^{\shortmid }\Im _{\ \ \beta _{s}}^{\star \alpha
_{s}}(\ ^{\widehat{a}}\xi )$ are chosen in a form (\ref{cosmsolitsourc}).

Let us provide some explicit formulas for nonlnear symmetries (\ref%
{nonlinsym}), which allow to change the data for generating functions \ and
generating sources from (\ref{solitdistr}), (\ref{nconc}) and (\ref%
{cosmsolitsourc}):

\begin{equation}
\ _{2}\Phi (\tau )=-4\ _{2}\Lambda (\tau )\eta _{3}(\tau )\mathring{g}_{3},\
_{3}\Phi (\tau )=-4\ _{3}^{\shortmid }\Lambda (\tau )\ ^{\shortmid }\eta
^{5}(\tau )\ ^{\shortmid }\mathring{g}^{5},\ _{4}\Phi (\tau )=-4\
_{4}^{\shortmid }\Lambda (\tau )\ ^{\shortmid }\eta ^{7}(\tau )\ ^{\shortmid
}\mathring{g}^{7}.\   \label{cosmologns}
\end{equation}%
For simplicity, we can chose such $\tau $-running horizontal shells
cosmological constants, $_{1}\Lambda (\tau )=\ _{2}\Lambda (\tau )=\Lambda
(\tau ),$ and co-vertical shells cosmological constants, $\ _{3}^{\shortmid
}\Lambda (\tau )=\ _{4}^{\shortmid }\Lambda (\tau )=\ ^{\shortmid }\Lambda
(\tau ).$ Such re-definition of generating data, $[\eta _{3},\ ^{\shortmid
}\eta ^{5},\ ^{\shortmid }\eta ^{7};~_{s}^{\shortmid }\mathcal{K}%
]\rightarrow \lbrack \ _{2}\Phi ,\ _{3}\Phi ,\ _{4}\Phi ;\Lambda ,\
^{\shortmid }\Lambda ]$ transform $\tau $-families of s-metrics $^{\shortmid
}\mathbf{g}(\tau )=\ ^{\shortmid }g[\ ^{\widehat{a}}\wp ,\ ^{\widehat{a}}\xi
]$ (\ref{cosmoffds}) into respective cosmological solutions of
nonassociative Ricci soliton equations (\ref{nariccisol}) when $\
_{1}\Lambda (\tau _{0})=\ _{2}\Lambda (\tau _{0})$ and $\ _{3}^{\shortmid
}\Lambda (\tau _{0})=\ _{4}^{\shortmid }\Lambda (\tau _{0}).$ In such cases,
we can use $\tau $-running, or with a fixed $\tau _{0},$ to model DE effects
in modern accelerating cosmology. \ Here we note that to study DM models and
respective cosmological solutions we should consider another type of
nonlinear transforms (\ref{cosmologns}) defined, for instance, in \cite%
{partner02,partner04}. Nonlinear DM distributions and structure formations
can be modelled for another type parameterizations of the effective matter
sources when $~_{s}^{\shortmid }\mathcal{K(\tau )}=\ _{s}^{m}\mathcal{K(\tau
)}+\ ^{DE,DM}\mathcal{K}[\ ^{\widehat{a}}\wp ,\ ^{\widehat{a}}\xi ]).$ For
such parameterizations, $\ _{s}^{m}\mathcal{K(\tau )}$ is defined by
observable matter fields, but $\ ^{DE,DM}\mathcal{K(\tau )}$ encode various
nonassociative/ noncommutative and another type contributions and geometric
distortions which may contribute to locally anisotropic DM and DE observable
cosmological evolution. Such details will be presented in our further
partner works. In this work, we analyze in brief only the cases with
cosmological geometric thermodynamics when nonlinear symmetries (\ref%
{cosmologns}) may transform be of type $~_{s}^{\shortmid }\mathcal{K}(\tau
)\rightarrow \lbrack \Lambda (\tau )=\ ^{m}\Lambda (\tau )+\ ^{DE}\Lambda
(\tau ),\ ^{\shortmid }\Lambda (\tau )],$ considering that co-fiber degrees
of freedom are approximated to $\ ^{\shortmid }\Lambda (\tau ).$
Nevertheless, we suppose that for 4-d spacetime projections, such (non)
associative off-diagonal interactions and evolution scenarious allow a
distinguishing into $\ ^{m}\Lambda (\tau )$ and $\ ^{DE}\Lambda (\tau ).$

\subsection{Perelman's thermodynamics for nonassociative cosmological
solutions}

\label{ssec32}

The class of nonassociative cosmological solitonic solutions (\ref{cosmoffds}%
) and their possible nonlinear transforms (\ref{cosmologns}) do not involve,
in general, any hypersurface, duality, or holographic configurations. This
means that the thermodynamics properties of such cosmological models can't
be described in the framework of the Bekenstein-Hawking paradigm \cite%
{bek2,haw2} but can be studied using a different type of G. Perelman's
thermodynamic variables \cite{perelman1}, see \cite{partner04,svnonh08} and
references therein for nonassociative and nonholonomic geometric flow models.

To compute the parametric R-flux deformed geometric flow thermodynamic
variable for nonassociative cosmological solitons (\ref{nariccisol}) we can
use formulas (61) from \cite{partner04} which for our prescription of
cosmological constants (see the end of previous section) at a fixed
temperature parameter $\tau _{0}$ are written in the form 
\begin{eqnarray}
\ _{s}^{\shortmid }\widehat{\mathcal{W}} &=&\int\nolimits_{\tau ^{\prime
}}^{\tau _{0}}\frac{d\tau }{16(\pi \tau )^{4}}\int_{\ ^{\shortmid }\widehat{%
\Xi }}\left( \tau \lbrack \ ^{m}\Lambda (\tau )+\ ^{DE}\Lambda (\tau )+\
^{\shortmid }\Lambda (\tau )]^{2}-2\right) \ ^{\shortmid }\delta \
^{\shortmid }\mathcal{V}(\tau ),  \label{thvcann} \\
\quad \ _{s}^{\shortmid }\widehat{\mathcal{Z}} &=&\exp \left[
\int\nolimits_{\tau ^{\prime }}^{\tau _{0}}\frac{d\tau }{(2\pi \tau )^{4}}%
\int_{\ ^{\shortmid }\widehat{\Xi }}\ ^{\shortmid }\delta \ ^{\shortmid }%
\mathcal{V}(\tau )\right]  \notag \\
\ _{s}^{\shortmid }\widehat{\mathcal{E}} &=&-\int\nolimits_{\tau ^{\prime
}}^{\tau _{0}}\frac{d\tau }{64\pi ^{4}\tau ^{2}}\int_{\ ^{\shortmid }%
\widehat{\Xi }}\left( [\ ^{m}\Lambda (\tau )+\ ^{DE}\Lambda (\tau )+\
^{\shortmid }\Lambda (\tau )]-\frac{1}{\tau }\right) \ ^{\shortmid }\delta \
^{\shortmid }\mathcal{V}(\tau ),  \notag \\
\ _{s}^{\shortmid }\widehat{\mathcal{S}} &=&-\int\nolimits_{\tau ^{\prime
}}^{\tau _{0}}\frac{d\tau }{16(\pi \tau )^{4}}\int_{\ ^{\shortmid }\widehat{%
\Xi }}\left( \tau \lbrack \ ^{m}\Lambda (\tau )+\ ^{DE}\Lambda (\tau )+\
^{\shortmid }\Lambda (\tau )]-2\right) \ ^{\shortmid }\delta \ ^{\shortmid }%
\mathcal{V}(\tau ).  \notag
\end{eqnarray}
The Perelman's W-entropy, $\ _{s}^{\shortmid }\widehat{\mathcal{W}},$
statistical thermodynamic function, $\ _{s}^{\shortmid }\widehat{\mathcal{Z}}%
,$ thermodynamic energy, $\ _{s}^{\shortmid }\widehat{\mathcal{E}},$ and
thermodynamic entropy, $\ _{s}^{\shortmid }\widehat{\mathcal{S}},$ in (\ref%
{thvcann}) are determined by integration on a closed phase space volume $\
^{\shortmid }\widehat{\Xi }(t)$ when the time like variables run from $t_{1}$
(initial time) to $t_{2}$ (time for observations); and when the valolume
element $\ ^{\shortmid }\delta \ ^{\shortmid }\mathcal{V}(\tau )$ is
computed using the coefficients of an off-diagonal cosmological solution (%
\ref{cosmoffds}). Details on such technical comuptations for
quasi-stationary solutions are provided with respect to formulas (60) in 
\cite{partner04}, when for cosmological configurations we should change $%
y^{3}\rightarrow y^{4}=t,$ and, respectively, for indices of geometric
objects, $4\rightarrow 3.$ The explicit formulas for such a volume
functional $\ ^{\shortmid }\delta \ ^{\shortmid }\mathcal{V}(\tau )$ depend
on the types of parameters and generating data we use in constructing our
cosmological solitionic hierarchies. They are different for elaborating
different structure formation for DM, with respective scales, non-Newtonian
dynamics, filament structure etc. For the purposes of this work, we conclude
that the geometric thermodynamic variables (\ref{thvcann}) \ depend in
explicit form on effective $\tau $-running effective cosmological constants $%
\ ^{m}\Lambda (\tau ),\ ^{DE}\Lambda (\tau )$ and$\ ^{\shortmid }\Lambda
(\tau ).$ We may chose certain values of $\ ^{DE}\Lambda (\tau _{0})$ which
correspond to observable data for DE. The contributions to DE coming from
standard matter fields, $\ ^{m}\Lambda (\tau ),$ and from phase (co) fibers, 
$\ ^{\shortmid }\Lambda (\tau ),$ also can be evaluated. For instance, $\
^{\shortmid }\Lambda (\tau )$ is effective determined by nonassociative
R-flux deformations even mixing of DE and DM effects are encoded into $\
^{\shortmid }\delta \ ^{\shortmid }\mathcal{V}(\tau ).$ This provides us
statistical and geometric thermodynamic picture of nonassociative cosmology
using the formalism of nonassociative and geometric flows for a
corresponding $\tau $-family of generic off-diagonal cosmological solutions.

\section{Conclusions and perspectives}

\label{sec4} Designing new perspectives in the modern cosmology of nonassociative and noncommutative theories, we formulated a model of gauge gravity constructed as a nonassociative R-flux generalization of the
noncommutative geometric constructions from \cite{svnc00}. For projections on the base spacetime, the commutative part of our theory is equivalent to GR, when the constructions on the total bundle spaces enabled with twisted star product can be parameterized and projected on phase spaces in certain forms which are equivalent to nonassocitative modifications of GR in \cite{blumenhagen16,aschieri17,partner02,partner04}. Corresponding modified Yang-Mills equations can be projected on a base spacetime (and on corresponding phase spaces) to result in the associative and commutative limits into standard Einstein equations. On generalization to suppersymmetric, noncommutative, nonholonomic generalizations, we cite 
\cite{vd00,svmp05,sv00}). This is different from other types of noncommutative theories 
\cite{utiyama,kibble,hehl,jurco,chamseddine,nicolini,chaichian,ciric} where more substantial gauge-like modifications of GR to MGTs with torsion and nonmetricity fields where considered. In our approach, torsion fields can be completely determined by nonholonomic effects (which can be used for decoupling in a general form of physically important systems of nonlinear PDEs) when all geometric constructions can be nonholonomically constrained to result in nonassociative/ noncommutative / commutative Levi-Civita
configurations.

\vskip4pt The new results of this paper are the following:

\begin{itemize}
\item We proved that the nonassociative gauge gravity can be constructed in such a form that it may preserve the de Sitter, dS, group, $SO(4,1)$, or affine group $Af(4,1)$, with the Poincar\'{e} group, $ISO(3,1)$, (with a corresponding dubbing), even the base Einstein manifolds and phase space are subjected to twisted $\star$-deformations. For R-fluxes, this encodes contributions from string theory. As it could be seen, the corrections to GR emerged with parametric dependence on the string and Planck constants, respectively, $\kappa$ and $\hbar .$ Such a theory is nonassociative and noncommutative even does not involve quantum groups, octonions etc.; locally, it depends on spacetime and momentum (or velocity) type coordinates.

\item Applying the AFCDM \cite{svmp05,sv15a,partner02,partner04}, we constructed a new class of generic off-diagonal cosmological solutions describing nonassociative and noncommutative deformations determined by generating functions and generating sources, and respective nonlinear symmetries. Such values can be chosen to define nonlinear solitonic/ wave hierarchies which may encode and distinguish nonassociative and noncommutative data and explain DE and DM effects, inflation and accelerating cosmological phases.

\item Also, we concluded that new classes of cosmological solutions in nonassociative gauge gravity theories can't be described in the framework of the Bekenstein-Hawking paradigm \cite{bek2,haw2} because, in general, they do not involve any hypersurface, duality or holographic conditions. Nevertheless, such solutions and respective DE and DM configurations can be characterized by a respective nonassociative generalized G. Perelman thermodynamics \cite{perelman1,svnonh08,kehagias19,partner04}. We show how to compute in explicit form respective W-entropy and thermodynamic variables encoding nonassociative and/or noncommutative data resulting in effective time and/or temperature running of the cosmological constant and other physical variables.
\end{itemize}

Finally, we mention three additional open problems that require further consideration (related to our research program on nonassociative geometric flows and gravity and applications in modern cosmology and quantum
information theories as we stated in \cite{sv15a,partner04}, see also their references to other partner works). First, we emphasize that the constructions with respective nonassociative star product deformed spinor
spaces and respective Einstein-Dirac equations have to be defined to result in a self-consistent viable model of nonassociative gauge gravity. Secondly, certain examples of exact/ parametric solutions for nonassociative
Einstein-Dirac-YM-Higgs systems have to be constructed, for instance, developing the AFCDM. And the third important problem is that such nonassociative and noncommutative theories, in hidden form, consist of examples of nonassociative/ noncommutative and complex Finsler-Lagrange-Hamilton spaces (because locally, the phase spaces enabled with twisted R-flux product depend also on momentum/velocity type
variables). It will be necessary to elaborate on such geometric and physical nonassociative gauge theories. Such problems are planned to be solved in our future work.

\vskip4pt \textbf{Acknowledgements:} This is a partner work of SV and co-authors for performing a research program on nonassociative geometric and quantum information flows and applications in modern gravity, see details in \cite{partner02,partner04}. 

\appendix

\setcounter{equation}{0} \renewcommand{\theequation}
{A.\arabic{equation}} \setcounter{subsection}{0} 
\renewcommand{\thesubsection}
{A.\arabic{subsection}}

\section{Nonassociative star products adapted to N-connection structures}

\label{appendixa}The articles \cite{partner02,partner04} provide reviews on
nonassociative geometric and quantum information flows and MGTs defined by
star products determined by R-flux deformations. In such works, the
nonassociative phase spaces are constructed as star deformations $\mathcal{M}%
\rightarrow \ \mathcal{M}^{\star }$ of some commutative phase spaces $%
\mathcal{M}=T\mathbf{V}$ (with local spacetime and velocity type
coordinates, $u=\{u^{\alpha }=(x^{i},v^{a})\})$, or $\ ^{\shortmid }\mathcal{%
M}=T^{\ast }\mathbf{V}$ (with spacetime and momentum like coordinates, $\
^{\shortmid }u=\{\ ^{\shortmid }u^{\alpha }=(x^{i},p_{a})\}).$ The total
spaces $T\mathbf{V}$ and $T^{\ast }\mathbf{V}$ are respective tangent and
cotangent bundles of a Lorentzian spacetime manifold $\mathbf{V}$ of
signature $(+++-).$ We follow an abstract (index and coordinate-free)
geometric formalism in GR \cite{misner}, which was generalized
correspondingly for research on nonassociative star-geometry in \cite%
{aschieri17} and, in nonholonomic form (adapted to nonlinear connection,
N-connection, structures), \cite{partner02}. For simplicity, we shall
provided only formulas for geometric and physical objects on $\ ^{\shortmid }%
\mathcal{M}$, when the constructions on $\mathcal{M}$ are similar (with a
formal omitting of the duality label "$\ ^{\shortmid }$", when in local
coordinates $p_{a}\rightarrow v^{a},$ when covertical, c, indices transforms
into resepctive vertical, v, ones).

The geometric constructions on $\ ^{\shortmid }\mathcal{M}$ can be adapted
to a N-connection structure, $\ ^{\shortmid }\mathbf{N}:TT^{\ast }\mathbf{V}%
=hT^{\ast }\mathbf{V}\oplus cT^{\ast }\mathbf{V}$, where $\oplus $ denotes a
Whithney direct sum. So, a N-connection $\ ^{\shortmid }\mathbf{N}$ is
defined as a nonholonomic (equivalently, anholonomic, or non-integrable)
distribution with conventional $(h,c)$-splitting of dimensions. In our
works, we use "bold face" symbols if it is important to note that the
geometric constructions are adapted to a N-connection splitting. In
N-adapted forms, tensors transforms into d-tensors, vectors transforms into
d-vectors and connections into d-connections, where "d" means distinguished
by N-connection h-c-splitting.

A distinguished connection, d-connection $\ ^{\shortmid }\mathbf{D=(}h\
^{\shortmid }D,c\ ^{\shortmid }D),$ preserves such a 4+4 spliting under
affine linear transports (as a typical linear connection). Here, we note
that a LC-connection $\ ^{\shortmid }\mathbf{\nabla }$ (which by definition
is metric compatible and torsionless) is not a d-connection because it is
not adapted to general N-connection structure. Nevertheless, we can always
define an N-adapted distortion formula$\ ^{\shortmid }\mathbf{D=\
^{\shortmid }\nabla +\ ^{\shortmid }Z,}$ where $\mathbf{^{\shortmid }Z}$ is
the distortion d-tensor encoding contributions from respective torsion of $\
^{\shortmid }\mathbf{D},$ and non-metricity d-tensor, $\mathbf{\ ^{\shortmid
}Q=:Dg,}$ when $\mathbf{\ ^{\shortmid }\nabla \ ^{\shortmid }g=0}$.Such
formulas can be written on a $\ \mathcal{M}$ with $\ \mathbf{D,}$ when $%
\mathbf{D=(}hD,vD),\mathbf{D=\nabla +Z,}$ $\mathbf{Q=:Dg,}$ and $\mathbf{%
\nabla g=0.}$

For simplicity, we shall prefer to work with the so-called canonical
d-connection $\ ^{\shortmid }\widehat{\mathbf{D}}\mathbf{,}$ which satisfy
the property that the canonical d-torsion tensor$\mathbf{\ ^{\shortmid }}%
\widehat{\mathcal{T}}=\{hh\mathbf{\ ^{\shortmid }}\widehat{\mathcal{T}}=0;cc%
\mathbf{\ ^{\shortmid }}\widehat{\mathcal{T}}=0,$ but $hc\mathbf{\
^{\shortmid }}\widehat{\mathcal{T}}\neq 0\}\neq 0$ is completely determined
by the coefficients of $\mathbf{\ ^{\shortmid }g}$ and $\mathbf{\
^{\shortmid }N}$ as a nonholonomic distortion effect. The geometric data $(%
\mathbf{\ ^{\shortmid }g,}\ ^{\shortmid }\widehat{\mathbf{D}})$ allow us to
prove certain general decoupling and integrability properties of physically
important systems of nonlinear PDEs using generic off-diagonal d-metrics $%
\mathbf{\ ^{\shortmid }g}$ (which can't be diagonalized by coordinate
transforms and may depend, in general, on all spacetime and phase space
coordinates). In our works, we omit "hat" on symbols or even the duality
label "$\mathbf{\ ^{\shortmid }}$" if such simplification of labels do not
result in ambiguities. Then, imposing additional nonholonomic constraints,
we shall be able to transform all classes of solutions into those for $(%
\mathbf{\ ^{\shortmid }g,}\ ^{\shortmid }\mathbf{\nabla }).$ Necessary
details on such (non) commutative nonholonomic geometric models and gravity
theories and the respective techniques for the anholonomic frame and
connction deformation method, AFCDM, can be found in \cite{vd00,svmp05,sv15a}
with further developments in nonassociative form \cite{partner02,partner04}.
The AFCDM allows to generate exact and parametric solutions using geometric
and analytic methods when the N-connection coefficients are non-trivial and
metrics can be generic off-diagonal.

To apply in explicit form the AFCDM for constructing exact/parametric
solutions encoding nonassociative data we have to follow also a nonholonomic
shell decomposition formalism, which allows us to decouple and integrate
various classes modified Einstein equations \cite{partner02}. In such cases,
a phase space $\ ^{\shortmid }\mathcal{M}$ is enabled with conventional
(2+2)+(2+2) splitting determined by a nonholonomic dyadic, 2-d,
decomposition into four oriented shells $s=1,2,3,4.$ In brief, we shall use
the term s-decomposition, when 
\begin{equation}
\ _{s}^{\shortmid }\mathbf{N}:\ \ _{s}T\mathbf{T}^{\ast }\mathbf{V}=\
^{1}hT^{\ast }V\oplus \ ^{2}vT^{\ast }V\oplus \ ^{3}cT^{\ast }V\oplus \
^{4}cT^{\ast }V,\mbox{  for }s=1,2,3,4,  \label{scon}
\end{equation}%
which in a local coordinate basis is characterized by a corresponding set of
coefficients $\ _{s}^{\shortmid }\mathbf{N}=\{\ ^{\shortmid }N_{\
i_{s}a_{s}}(\ ^{\shortmid }u)\},$ for any point $u=(x,p)=\ ^{\shortmid }u=(\
_{1}x,\ _{2}y,\ _{3}p,\ _{4}p)\in \mathbf{T}^{\ast }\mathbf{V}$. This allows
us to introduce in s-coefficient form some N-elongated bases (N-/ s-adapted
bases as linear N-operators): 
\begin{eqnarray}
\ ^{\shortmid }\mathbf{e}_{\alpha _{s}}[\ ^{\shortmid }N_{\ i_{s}a_{s}}]
&=&(\ ^{\shortmid }\mathbf{e}_{i_{s}}=\ \frac{\partial }{\partial x^{i_{s}}}%
-\ ^{\shortmid }N_{\ i_{s}a_{s}}\frac{\partial }{\partial p_{a_{s}}},\ \
^{\shortmid }e^{b_{s}}=\frac{\partial }{\partial p_{b_{s}}})\mbox{ on }\
_{s}T\mathbf{T}_{\shortmid }^{\ast }\mathbf{V;}  \notag \\
\ ^{\shortmid }\mathbf{e}^{\alpha _{s}}[\ ^{\shortmid }N_{\ i_{s}a_{s}}]
&=&(\ ^{\shortmid }\mathbf{e}^{i_{s}}=dx^{i_{s}},\ ^{\shortmid }\mathbf{e}%
_{a_{s}}=d\ p_{a_{s}}+\ ^{\shortmid }N_{\ i_{s}a_{s}}dx^{i_{s}})\mbox{ on }\
\ _{s}T^{\ast }\mathbf{T}_{\shortmid }^{\ast }\mathbf{V.}  \label{nadapb}
\end{eqnarray}%
We put a left label $s$ for corresponding spaces and geometric objects if,
for instance, it is necessary to emphasize that a phase space is enabled
with a s-adapted dyadic structure denoting $\ _{s}^{\shortmid }\mathcal{M}$.
Having prescribed a dyadic s-structure, we can express any metric or
d-metric as a s-metric $\ _{s}^{\shortmid }\mathbf{g}=\{\ ^{\shortmid }%
\mathbf{g}_{\alpha _{s}\beta _{s}}\},$ when the respective s-tensor
components, i.e. s-adapted coefficients, are parameterized $\ ^{\shortmid
}g=\ _{s}^{\shortmid }\mathbf{g}=(h_{1}\ ^{\shortmid }\mathbf{g},~v_{2}\
^{\shortmid }\mathbf{g},\ c_{3}\ ^{\shortmid }\mathbf{g,}c_{4}\ ^{\shortmid }%
\mathbf{g})\in T\mathbf{T}^{\ast }\mathbf{V}\otimes T\mathbf{T}^{\ast }%
\mathbf{V,}$ and $\ ^{\shortmid }g=\ _{s}^{\shortmid }\mathbf{g}=\
^{\shortmid }\mathbf{g}_{\alpha _{s}\beta _{s}}(\ _{s}^{\shortmid }u)\ \
^{\shortmid }\mathbf{e}^{\alpha _{s}}\otimes _{s}\ ^{\shortmid }\mathbf{e}%
^{\beta _{s}}=\{\ \ ^{\shortmid }\mathbf{g}_{\alpha _{s}\beta _{s}}=(\ \
^{\shortmid }\mathbf{g}_{i_{1}j_{1}},\ \ ^{\shortmid }\mathbf{g}%
_{a_{2}b_{2}},\ \ ^{\shortmid }\mathbf{g}^{a_{3}b_{3}},\ \ ^{\shortmid }%
\mathbf{g}^{a_{4}b_{4}})\},$where $\ ^{\shortmid }\mathbf{e}^{\alpha _{s}}$ (%
\ref{nadapb}) can be chosen in s-adapted form.

In our approach \cite{partner02,partner04} (generalizing in nonholonomic
s-adapted form the constructions from \cite{blumenhagen16,aschieri17}),
nonassociative and noncommutative geometric and gravity theories are defined
by a twisted star product involvig actions of N-elongated differential
operators $\ ^{\shortmid }\mathbf{e}_{i_{s}}$ (\ref{nadapb}), on some
functions $\ f(x,p)$ and $\ q(x,p)$ defined on a phase space $\ \
_{s}^{\shortmid }\mathcal{M}$. For such a s-adapted star product $\star
_{s}, $ we can compute always 
\begin{eqnarray}
f\star _{s}q:= &&\cdot \lbrack \mathcal{F}_{s}^{-1}(f,q)]  \label{starpn} \\
&=&\cdot \lbrack \exp (-\frac{1}{2}i\hbar (\ ^{\shortmid }\mathbf{e}%
_{i_{s}}\otimes \ ^{\shortmid }e^{i_{s}}-\ ^{\shortmid }e^{i_{s}}\otimes \
^{\shortmid }\mathbf{e}_{i_{s}})+\frac{i\mathit{\ell }_{s}^{4}}{12\hbar }%
R^{i_{s}j_{s}a_{s}}(p_{a_{s}}\ ^{\shortmid }\mathbf{e}_{i_{s}}\otimes \
^{\shortmid }\mathbf{e}_{j_{a}}-\ ^{\shortmid }\mathbf{e}_{j_{s}}\otimes
p_{a_{s}}\ ^{\shortmid }\mathbf{e}_{i_{s}}))]f\otimes q  \notag \\
&=&f\cdot q-\frac{i}{2}\hbar \lbrack (\ ^{\shortmid }\mathbf{e}_{i_{s}}f)(\
^{\shortmid }e^{i_{s}}q)-(\ ^{\shortmid }e^{i_{s}}f)(\ ^{\shortmid }\mathbf{e%
}_{i_{s}}q)]+\frac{i\mathit{\ell }_{s}^{4}}{6\hbar }%
R^{i_{s}j_{s}a_{s}}p_{a_{s}}(\ ^{\shortmid }\mathbf{e}_{i_{s}}f)(\
^{\shortmid }\mathbf{e}_{j_{s}}q)+\ldots ..  \notag
\end{eqnarray}%
This twisted by antisymmetric coefficients $R^{i_{s}j_{s}a_{s}}$ star
product (in brief, $\star $ and, correspondingly $\star _{N},$ or $\star
_{s},$ the constant $\mathit{\ell }$ characterizes the R-flux contributions
from string theory. Here we note that the tensor product $\otimes $ can be
written also in a s-adapted form $\otimes _{s}.$ This way, we can construct
star product deformed gravity and field theories, when explicit computations
for R-flux deformations of s-adapted geometric objects and (physical)
equations, cand be adapted and classified with respect to decompositions on
small parameters $\hbar $ and $\kappa =\mathit{\ell }_{s}^{3}/6\hbar .$ In
such cases, all tensor products turn into usual multiplications as in the
third line of above formula.

A star product (\ref{starpn}) transforms a $\ _{s}^{\shortmid }\mathcal{M}$
(and any geometric s-objects for a (co) s-vector bundle $\ _{s}^{\shortmid }%
\mathcal{E}(\ _{s}^{\shortmid }\mathcal{M})$ and base $\ _{s}^{\shortmid }%
\mathcal{M}$) into respecive nonassociative phase spaces, their \ bundle
spaces and s-objects, denoted in abstract form, for instance, $\ \
_{s}^{\shortmid }\mathcal{M}^{\star },$ $\ _{s}^{\shortmid }\mathcal{E}%
^{\star }$ etc. In abstract form, such a techniques works both for
commutative and nonassociative/ noncommutative spaces. For instance, in \cite%
{svnc00,sv00}, we used a similar N-adapted Siberg-Witten star product $\ast
, $ or $\ast _{N}$. Nevertheless, it is not just a formal changing of, for
instance, $\ast _{N}$into a $\star _{s}$ (\ref{starpn}) without R-flux
terms. The main issue is that R-flux $\star $-deformations for the metrics, $%
\star :\mathbf{g\rightarrow g}^{\star }=(\mathbf{\breve{g}}^{\star },\mathbf{%
\check{g}}^{\star }),$ result in certain nonassociative symmetric, $\breve{g}%
^{\star }$, and nonassociative nonsymmetric, $\mathbf{\check{g}}^{\star }$,
components, when the non-symmetry of metrics may be avoided in
noncommutative $\ast $-theories.

Abstract geometric and/or tedious index/ coordinate computations of the main
geometric and physical objects on $\ _{s}^{\shortmid }\mathcal{M}^{\star }$%
allow us to express all important formulas for the "star" -d-metrics,
d-connections, d-torsions, d-curvatures etc. into certain $\hbar $ and $%
\kappa $-parametric forms "without stars" as in \cite{aschieri17,partner02}.
Such computations can be considered for the $\star $-versions of
LC-connections, $\mathbf{\nabla \rightarrow \nabla }^{\star };$ arbitrary
d-connections, $\ ^{\shortmid }\mathbf{D\rightarrow }\ ^{\shortmid }\mathbf{D%
}^{\star },$ or canonical s-connections,$\ _{s}^{\shortmid }\widehat{\mathbf{%
D}}\mathbf{\rightarrow }\ _{s}^{\shortmid }\widehat{\mathbf{D}}^{\star },$
etc. Correspondingly, we can compute the parametric and s-adapted forms for
a star product deformation of the Ricci tensor, or canonical s-tensor, $%
\mathcal{R}ic^{\star }[\mathbf{g}^{\star },\mathbf{\nabla }^{\star }]$ or $%
\widehat{\mathcal{R}}ic^{\star }[\mathbf{g}^{\star },\widehat{\mathbf{D}}%
^{\star }]$ etc. The $\hbar $ and $\kappa $-parametric terms determined by $%
\star $ deformations of pseudo-Riemannian metrics can be re-defined
equivalently as certain effective sources encoding nonassociative/
noncommutative data from string theory, see Convention 2 and related details
in \cite{partner02,partner04}.

The coordinates on the nonassociatve phase spaces are parameterized as in
Appendix A.1 to \cite{partner04} when 
\begin{eqnarray}
&\mbox{on}&T_{\shortparallel }^{\ast }\mathbf{V}\mbox{ and }%
T_{\shortparallel s}^{\ast }\mathbf{V}:\ ^{\shortparallel }u=(x,\
^{\shortparallel }p)=\{\ ^{\shortparallel }u^{\alpha }=(u^{k}=x^{k},\
^{\shortparallel }p_{a}=(i\hbar )^{-1}p_{a})\}  \notag \\
&=&(\ _{3}^{\shortparallel }x,\ _{4}^{\shortparallel }p)=\{\
^{\shortparallel }u^{\alpha }=(^{\shortparallel }u^{k_{3}}=\
^{\shortparallel }x^{k_{3}},\ ^{\shortparallel }p_{a_{4}}=(i\hbar
)^{-1}p_{a_{4}})\}=  \label{coordconv} \\
&&\ _{s}^{\shortparallel }u=(\ _{s}x,\ _{s}^{\shortparallel }p)=\{\
^{\shortparallel }u^{\alpha _{s}}=(x^{k_{s}},\ ^{\shortparallel
}p_{a_{s}}=(i\hbar )^{-1}p_{a_{s}})\}=(x^{i_{1}},x^{i_{2}},\
^{\shortparallel }p_{a_{3}}=(i\hbar )^{-1}p_{a_{3}},\ ^{\shortparallel
}p_{a_{4}}=(i\hbar )^{-1}p_{a_{4}}),  \notag \\
&=&(\ _{3}^{\shortparallel }u~=\ _{3}^{\shortparallel }x,\
_{4}^{\shortparallel }p)=\{\ ^{\shortparallel }u^{\alpha
_{3}}=(x^{i_{1}},x^{i_{2}},\ ^{\shortparallel }x^{i_{3}}\rightarrow \
^{\shortparallel }p_{a_{3}}),\ ^{\shortparallel }p_{a_{4}}\},\mbox{ where }\
^{\shortparallel }x^{\alpha _{3}}=(x^{i_{1}},x^{i_{2}},\ ^{\shortparallel
}p_{a_{3}}=(i\hbar )^{-1}p_{a_{3}}).  \notag
\end{eqnarray}%
In these formulas, the coordinate $x^{4}=y^{4}=t$ is time-like and $p_{8}=E$
is energy-like. We consider boldface indices spaces and geometric objects
enabled with N-/s-connection structure $\ _{s}^{\shortparallel }\mathbf{N}$
depending coordinates $_{s}^{\shortparallel }u$. An upper or lower left
label "$\ ^{\shortparallel }$" is used to distinguish coordinates with
\textquotedblleft complexified momenta" of real phase coordinates $\
^{\shortmid }u^{\alpha }=(x^{k},p_{a})$ on $T^{\ast }\mathbf{V}$.

Tedious parametric computations with separation of coefficients proportional
to $\hbar ,\kappa $ and $\hbar \kappa ,$ see \cite%
{aschieri17,partner02,partner04} allow to express the nonassociative
gravitational field equations (\ref{nonassocaneinst}) in the form 
\begin{eqnarray}
\ ^{\shortparallel }\widehat{\mathbf{R}}_{\ \beta _{s}\gamma _{s}}~ &=&\
^{\shortparallel }\mathbf{K}_{_{\beta _{s}\gamma _{s}}},%
\mbox{ for effective
nonassociative sources }  \label{cannonsymparamc2} \\
\ ^{\shortparallel }\mathbf{K}_{_{\beta _{s}\gamma _{s}}} &=&\
_{[0]}^{\shortparallel }\Upsilon _{_{\beta _{s}\gamma _{s}}}+\
_{[1]}^{\shortparallel }\mathbf{K}_{_{\beta _{s}\gamma _{s}}}\left\lceil
\hbar ,\kappa \right\rceil ,\mbox{ where }\ _{[0]}^{\shortparallel }\Upsilon
_{_{\beta _{s}\gamma _{s}}}=\ ^{s}\Lambda (\ ^{\shortparallel }u^{\gamma
_{s}})\ _{\star }^{\shortparallel }\mathbf{g}_{\beta _{s}\gamma _{s}}\ %
\mbox{and }  \notag \\
&&\ _{[1]}^{\shortparallel }\mathbf{K}_{_{\beta _{s}\gamma _{s}}}\left\lceil
\hbar ,\kappa \right\rceil =\ ^{s}\Lambda (\ ^{\shortparallel }u^{\gamma
_{s}})\ _{\star }^{\shortparallel }\mathbf{\check{q}}_{\beta _{s}\gamma
_{s}}^{[1]}(\kappa )-\ ^{\shortparallel }\widehat{\mathbf{K}}_{\ \beta
_{s}\gamma _{s}}\left\lceil \hbar ,\kappa \right\rceil .  \notag
\end{eqnarray}%
The effective sources (\ref{cannonsymparamc2}) can be parameterized for
nontrivial real cosmological 8-d phase space configurations using
coordinates $(x^{k_{3}},\ ^{\shortparallel }p_{8}),$ for $\ ^{\shortmid
}p_{8}=E,$ with $\ _{\star }^{\shortparallel }\mathbf{g}_{\beta _{s}\gamma
_{s}\mid \hbar ,\kappa =0}=\ ^{\shortparallel }\mathbf{g}_{\beta _{s}\gamma
_{s}},$ $x^{4}=t,$ in such forms: {\small 
\begin{eqnarray*}
\ ^{\shortparallel }\mathbf{K}_{\ \beta _{s}}^{\alpha _{s}} &=&\{\
^{\shortparallel }\mathcal{K}_{\ j_{1}}^{i_{1}}(\kappa ,x^{k_{1}})=[\ \
_{1}^{\shortparallel }\Upsilon (x^{k_{1}})+\ _{1}^{\shortparallel }\mathbf{K}%
(\kappa ,x^{k_{1}})]\delta _{j_{1}}^{i_{1}},\ ^{\shortparallel }\mathcal{K}%
_{\ j_{2}}^{i_{2}}(\kappa ,x^{k_{1}},t)=[\ _{2}^{\shortparallel }\Upsilon
(x^{k_{1}},t)+\ _{2}^{\shortparallel }\mathbf{K}(\kappa ,x^{k_{1}},t)]\delta
_{b_{2}}^{a_{2}}, \\
&&\ ^{\shortparallel }\mathcal{K}_{\ a_{3}}^{b_{3}}(\kappa ,x^{k_{2}},\
^{\shortparallel }p_{6})=[\ \ _{3}^{\shortparallel }\Upsilon (x^{k_{2}},\
^{\shortparallel }p_{6})+\ _{3}^{\shortparallel }\mathbf{K}(x^{k_{2}},\
^{\shortparallel }p_{6})]\ \delta _{a_{3}}^{b_{3}}, \\
&&\ ^{\shortparallel }\mathcal{K}_{\ a_{4}}^{b_{4}}(\kappa ,x^{k_{3}},\
^{\shortparallel }p_{8})=[\ \ _{4}^{\shortparallel }\Upsilon (x^{k_{3}},\
^{\shortparallel }p_{8})+\ _{4}^{\shortparallel }\mathbf{K}(x^{k_{3}},\
^{\shortparallel }p_{8})]\delta _{a_{4}}^{b_{4}}\},\mbox{ where }\
^{\shortparallel }\mathbf{K}_{\ j_{s}k_{s}}=-\ _{[11]}^{\shortparallel }%
\widehat{\mathbf{R}}ic_{j_{s}k_{s}}^{\star }(\ ^{\shortparallel
}u^{k_{s-2}},\ ^{\shortparallel }u^{k_{s}}), \\
\mathbf{g}_{j_{s}k_{s}}
&=&%
\{g_{1}(x^{k_{1}}),g_{2}(x^{k_{1}}),g_{3}(x^{k_{1}},x^{3}),g_{4}(x^{k_{1}},x^{3}),\ ^{\shortparallel }g^{5}(x^{k_{2}},\ ^{\shortparallel }p_{6}),\ ^{\shortparallel }g^{6}(x^{k_{2}},\ ^{\shortparallel }p_{6}),\ ^{\shortparallel }g^{7}(x^{k_{3}},\ ^{\shortparallel }p_{8}),\ ^{\shortparallel }g^{8}(x^{k_{3}},\ ^{\shortparallel }p_{8})\}.
\end{eqnarray*}%
} Considering frame transforms $\ ^{\shortparallel }\Im _{\alpha
_{s}^{\prime }\beta _{s}^{\prime }}=e_{\ \alpha _{s}^{\prime }}^{\alpha
_{s}}e_{\ \beta _{s}^{\prime }}^{\beta _{s}}\ \ ^{\shortparallel }\mathcal{K}%
_{\alpha _{s}\beta _{s}}$, we parameterize 
\begin{equation}
\ ^{\shortmid }\Im _{\ \ \beta _{s}}^{\star \alpha _{s}}=\ ^{\shortparallel }%
\mathbf{K}_{\ \beta _{s}}^{\alpha _{s}}~=[~_{1}^{\shortparallel }\mathcal{K}%
(\kappa ,x^{k_{1}})\delta _{i_{1}}^{j_{1}},~_{2}^{\shortparallel }\mathcal{K}%
(\kappa ,x^{k_{1}},t)\delta _{b_{2}}^{a_{2}},~_{3}^{\shortparallel }\mathcal{%
K}(\kappa ,x^{k_{2}},\ ^{\shortparallel }p_{6})\delta
_{a_{3}}^{b_{3}},~_{4}^{\shortparallel }\mathcal{K}(\kappa ,x^{k_{3}},\
^{\shortparallel }p_{8})\delta _{a_{4}}^{b_{4}}].  \label{cannonsymparamc2a}
\end{equation}
For such parameterizations of sources, we can prove general decoupling and
integration properties of physically important systems of nonlinear PDEs in
nonassociative geometric and gravity theories. 

\setcounter{equation}{0} \renewcommand{\theequation}
{B.\arabic{equation}} \setcounter{subsection}{0} 
\renewcommand{\thesubsection}
{B.\arabic{subsection}}

\section{Cosmological solitonic hierarchies from generating functions/
sources}

\label{appendixb} In this appendix, we summarize necessary concepts and
formulas which are necessary for generating cosmological solitonic
hierarchies defined phase space s-metrics of type $\ ^{\shortmid }\mathbf{g}%
(\tau )$ (\ref{cosmoffds}).

Let us consider a non--stretching curve $\gamma (\tau ,\mathbf{l})$ on a
nonholonomic phase space $\ ^{\shortmid }\mathcal{M}=T^{\ast }\mathbf{V}$
when, for simplicity, a real $\tau $ is used both as a curve running real
parameter and a geometric flow parameter. We denote by $\mathbf{l}$ the
arclength of such a curve defined by an evolution d--vector $\ ^{\shortmid }%
\mathbf{Y}=\varsigma _{\tau }$ and tangent d--vector $\ ^{\shortmid }\mathbf{%
X}=\varsigma _{\mathbf{l}},$ for which $\ ^{\shortmid }\mathbf{g(\
^{\shortmid }X,\ ^{\shortmid }X)=}1$. Any $\varsigma (\tau ,\mathbf{l})$
defines a two--dimensional surface in $T_{\varsigma (\tau ,\mathbf{l})}\
^{\shortmid }\mathcal{M}\subset T\ ^{\shortmid }\mathcal{M}\mathbf{.}$ We
cite \cite{sv15a,partner02} and references therein for details on geometric
methods and cosmological applications of the theory of metric compatible
curve flows and solitonic hierarchies. Similar constructions are used in the
main part of this work in order to study nonassociative R-flux deformations
resulting in cosmological solitons.

We can associate a coframe $\ ^{\shortmid }\mathbf{e}\in T_{\varsigma}^{\ast
}\ ^{\shortmid }\mathcal{M}\otimes (h\mathfrak{p\oplus }v\mathfrak{p})$ to
any dual basis $\ ^{\shortmid }\mathbf{e}^{\alpha _{s}}$ (\ref{nadapb}),
considering respective $h$- and $v$-associated Lie algebras, $h\mathfrak{p}$
and $v\mathfrak{p},$ defining a s-adapted $\left( SO(n)\mathfrak{\oplus }%
SO(m)\right) $--parallel basis along $\varsigma .$ For geometric
constructions on such phase spaces, $n+m=8,$ we can consider $n=4$. But to
generate base spacetime solitonic hierarchies encoding contributions of
momentum like variables, we can fix any $n=2,3,5$ (for respective $m=6,5,3).$
Then, using a canonical d-connection $\ ^{\shortmid }\widehat{\mathbf{D}},$
we can define a linear d-connection 1--form parameterized as $\ ^{\shortmid }%
\widehat{\mathbf{\Gamma }}\in T_{\varsigma }^{\ast }\ ^{\shortmid }\mathcal{M%
} \otimes (\mathfrak{so}(n)\mathfrak{\oplus so}(m)).$ Respectively, using
the s-adapted frame bases to define 1-forms $\ ^{\shortmid }\mathbf{e}_{%
\mathbf{X}}=\ ^{\shortmid }\mathbf{e}_{h\mathbf{X}}+\ ^{\shortmid }\mathbf{e}%
_{v\mathbf{X}},$ and considering $(1,\overrightarrow{0})\in \mathbb{R}^{n},%
\overrightarrow{0}\in \mathbb{R}^{n-1}$ and $(1,\overleftarrow{0})\in 
\mathbb{R}^{m},\overleftarrow{0}\in \mathbb{R}^{m-1}),$ we define the
matrices: $\ ^{\shortmid }\mathbf{e}_{h\mathbf{X}}=\varsigma _{h\mathbf{X}%
}\rfloor h\ ^{\shortmid }\mathbf{e=}\left[ 
\begin{array}{cc}
0 & (1,\overrightarrow{0}) \\ 
-(1,\overrightarrow{0})^{T} & h\mathbf{0}%
\end{array}%
\right] $ and$\ ^{\shortmid }\mathbf{e}_{v\mathbf{X}}=\varsigma _{v\mathbf{X}%
}\rfloor v\ ^{\shortmid }\mathbf{e=}\left[ 
\begin{array}{cc}
0 & (1,\overleftarrow{0}) \\ 
-(1,\overleftarrow{0})^{T} & v\mathbf{0}%
\end{array}%
\right] .$ Such d-operators can be adapted also to any nonholonomic dyadic
variables and act on the spaces of curves on $\ ^{\shortmid }\mathcal{M}$.

For a N-connection (\ref{scon}) (considering double nonholonomic splittings
on phase space), the canonical d-connection 1-forms, $\ ^{\shortmid }%
\widehat{\mathbf{\Gamma }}=\left[ \ ^{\shortmid }\widehat{\mathbf{\Gamma }}%
_{h\mathbf{X}},\ ^{\shortmid }\widehat{\mathbf{\Gamma }}_{v\mathbf{X}}\right]
\in \lbrack \mathfrak{so}(n+1),\mathfrak{so}(m+1)],$ we can parameterize 
\begin{eqnarray*}
\ ^{\shortmid }\widehat{\mathbf{\Gamma }}_{h\mathbf{X}} &=&\varsigma _{h%
\mathbf{X}}\rfloor \ ^{\shortmid }\widehat{\mathbf{L}}=\left[ 
\begin{array}{cc}
0 & (0,\overrightarrow{0}) \\ 
-(0,\overrightarrow{0})^{T} & \ ^{\shortmid }\widehat{\mathbf{L}}%
\end{array}%
\right] ,\ ^{\shortmid }\widehat{\mathbf{\Gamma }}_{v\mathbf{X}}=\varsigma
_{v\mathbf{X}}\rfloor \ ^{\shortmid }\widehat{\mathbf{C}}=\left[ 
\begin{array}{cc}
0 & (0,\overleftarrow{0}) \\ 
-(0,\overleftarrow{0})^{T} & \ ^{\shortmid }\widehat{\mathbf{C}}%
\end{array}%
\right] ,\mbox{ where } \\
\ ^{\shortmid }\widehat{\mathbf{L}} &=&\left[ 
\begin{array}{cc}
0 & \overrightarrow{v} \\ 
-\overrightarrow{v}^{T} & h\mathbf{0}%
\end{array}%
\right] \in \mathfrak{so}(n),~\overrightarrow{v}\in \mathbb{R}^{n-1},~h%
\mathbf{0\in }\mathfrak{so}(n-1),\mbox{ and } \\
\ ^{\shortmid }\widehat{\mathbf{C}} &=&\left[ 
\begin{array}{cc}
0 & \overleftarrow{v} \\ 
-\overleftarrow{v}^{T} & v\mathbf{0}%
\end{array}%
\right] \in \mathfrak{so}(m),~\overleftarrow{v}\in \mathbb{R}^{m-1},~v%
\mathbf{0\in }\mathfrak{so}(m-1).
\end{eqnarray*}

A canonical s--connection $\ _{s}^{\shortmid }\widehat{\mathbf{D}}$ allows
also to define certain d-matrices being decomposed with respect to the h-
and v-splitting of the flow directions. For the h--direction, 
\begin{eqnarray*}
\ ^{\shortmid }\mathbf{e}_{h\mathbf{Y}} &=&\varsigma _{\tau }\rfloor h\
^{\shortmid }\mathbf{e}=\left[ 
\begin{array}{cc}
0 & \left( h\ ^{\shortmid }\mathbf{e}_{\parallel },h\ ^{\shortmid }%
\overrightarrow{\mathbf{e}}_{\perp }\right) \\ 
-\left( h\ ^{\shortmid }\mathbf{e}_{\parallel },h\ ^{\shortmid }%
\overrightarrow{\mathbf{e}}_{\perp }\right) ^{T} & h\mathbf{0}%
\end{array}%
\right] ,\mbox{ when } \\
\ ^{\shortmid }\mathbf{e}_{h\mathbf{Y}} &\in &h\mathfrak{p,}\left( h\
^{\shortmid }\mathbf{e}_{\parallel },h\ ^{\shortmid }\overrightarrow{\mathbf{%
e}}_{\perp }\right) \in \mathbb{R}^{n}\mbox{ and }h\ ^{\shortmid }%
\overrightarrow{\mathbf{e}}_{\perp }\in \mathbb{R}^{n-1},\mbox{ and }
\end{eqnarray*}
\begin{eqnarray*}
\ ^{\shortmid }\widehat{\mathbf{\Gamma }}_{h\mathbf{Y}} &=&\varsigma _{h%
\mathbf{Y}}\rfloor \ ^{\shortmid }\widehat{\mathbf{L}}=\left[ 
\begin{array}{cc}
0 & (0,\overrightarrow{0}) \\ 
-(0,\overrightarrow{0})^{T} & h\ ^{\shortmid }\mathbf{\varpi }_{\tau }%
\end{array}%
\right] \in \mathfrak{so}(n+1),\mbox{ where } \\
\ h\ ^{\shortmid }\mathbf{\varpi }_{\tau } &=&\left[ 
\begin{array}{cc}
0 & \ ^{\shortmid }\overrightarrow{\varpi } \\ 
-\ ^{\shortmid }\overrightarrow{\varpi }^{T} & h\ ^{\shortmid }\widehat{%
\mathbf{\Theta }}%
\end{array}%
\right] \in \mathfrak{so}(n),~\ ^{\shortmid }\overrightarrow{\varpi }\in 
\mathbb{R}^{n-1},~h\ ^{\shortmid }\widehat{\mathbf{\Theta }}\mathbf{\in }%
\mathfrak{so}(n-1).
\end{eqnarray*}%
Similar parameterizations can be defined for the v--direction: 
\begin{eqnarray*}
\ ^{\shortmid }\mathbf{e}_{v\mathbf{Y}} &=&\varsigma _{\tau }\rfloor v\
^{\shortmid }\mathbf{e}=\left[ 
\begin{array}{cc}
0 & \left( v\ ^{\shortmid }\mathbf{e}_{\parallel },v\ ^{\shortmid }%
\overleftarrow{\mathbf{e}}_{\perp }\right) \\ 
-\left( v\ ^{\shortmid }\mathbf{e}_{\parallel },v\ ^{\shortmid }%
\overleftarrow{\mathbf{e}}_{\perp }\right) ^{T} & v\mathbf{0}%
\end{array}%
\right] ,\mbox{ when } \\
\ ^{\shortmid }\mathbf{e}_{v\mathbf{Y}} &\in &v\mathfrak{p,}\left( v\
^{\shortmid }\mathbf{e}_{\parallel },v\ ^{\shortmid }\overleftarrow{\mathbf{e%
}}_{\perp }\right) \in \mathbb{R}^{m}\mbox{ and }v\ ^{\shortmid }%
\overleftarrow{\mathbf{e}}_{\perp }\in \mathbb{R}^{m-1};\mbox{ and }
\end{eqnarray*}
\begin{eqnarray*}
\ ^{\shortmid }\widehat{{\mathbf{\Gamma }}}_{v\mathbf{Y}} &=&\varsigma _{v%
\mathbf{Y}}\rfloor \ ^{\shortmid }\widehat{\mathbf{C}}= \left[ 
\begin{array}{cc}
0 & (0,\overleftarrow{0}) \\ 
-(0,\overleftarrow{0})^{T} & v\ ^{\shortmid }\widehat{\mathbf{\varpi }}%
_{\tau }%
\end{array}%
\right] \in \mathfrak{so}(m+1), \\
v\ ^{\shortmid }\mathbf{\varpi }_{\tau } &\mathbf{=}&\left[ 
\begin{array}{cc}
0 & \ ^{\shortmid }\overleftarrow{\varpi } \\ 
-\ ^{\shortmid }\overleftarrow{\varpi }^{T} & v\ ^{\shortmid }\widehat{%
\mathbf{\Theta }}%
\end{array}%
\right] \in \mathfrak{so}(m),~\ ^{\shortmid }\overleftarrow{\varpi }\in 
\mathbb{R}^{m-1},~v\ ^{\shortmid }\widehat{\mathbf{\Theta }}\mathbf{\in }%
\mathfrak{so}(m-1).
\end{eqnarray*}

Using above s-operators, we generalize for phase spaces a series of
important results. For any solution of the (gauge) gravitational equations (%
\ref{nonassocaneinst}) and/or (\ref{nonassocymgreq1}), there is a canonical
hierarchy of s--adapted flows of curves $\varsigma (\tau ,\mathbf{l}%
)=h\varsigma (\tau ,\mathbf{l})+v\varsigma (\tau ,\mathbf{l})$ described by
nonholonomic geometric map equations encoding nonassociative and
noncommutative sources:

\begin{enumerate}
\item The $0$ flows are defined by convective (also called travelling wave)
maps $\varsigma _{\tau }=\varsigma _{\mathbf{l}}$ distinguished as $\left(
h\varsigma \right) _{\tau }=\left( h\varsigma \right) _{h\mathbf{X}}$ and $%
\left( v\varsigma \right) _{\tau }=\left( v\varsigma \right) _{v\mathbf{X}}$%
. The classification of such maps depends on the type of d-connection
structure and nonholonomic dyadic splitting.

\item The +1 flows are defined as solutions for so-called non--stretching
mKdV map equations 
\begin{equation*}
-\left( h\varsigma \right) _{\tau }=\ ^{\shortmid }\widehat{\mathbf{D}}_{h%
\mathbf{X}}^{2}\left( h\varsigma \right) _{h\mathbf{X}}+\frac{3}{2}|\
^{\shortmid }\widehat{\mathbf{D}}_{h\mathbf{X}}\left( h\varsigma \right) _{h%
\mathbf{X}}|_{h\mathbf{g}}^{2}\left( h\varsigma \right) _{h\mathbf{X}%
},-\left( v\varsigma \right) _{\tau }=\ ^{\shortmid }\widehat{\mathbf{D}}_{v%
\mathbf{X}}^{2}\left( v\varsigma \right) _{v\mathbf{X}}+\frac{3}{2}|\
^{\shortmid }\widehat{\mathbf{D}}_{v\mathbf{X}}\left( v\varsigma \right) _{v%
\mathbf{X}}|_{v\mathbf{g}}^{2}~\left( v\varsigma \right) _{v\mathbf{X}},
\end{equation*}%
when the +2,... flows can be defined as higher order analogs.

\item Finally, the -1 flows are defined by the kernels of the canonical
recursion h--operator and, respective, v-operator, 
\begin{eqnarray*}
h\ ^{\shortmid }\widehat{{\mathfrak{R}}} &=&\ ^{\shortmid }\widehat{\mathbf{D%
}}_{h\mathbf{X}}\left( \ ^{\shortmid }\widehat{\mathbf{D}}_{h\mathbf{X}}+\
^{\shortmid }\widehat{\mathbf{D}}_{h\mathbf{X}}^{-1}\left( \ ^{\shortmid }%
\overrightarrow{v}\cdot \right) \ ^{\shortmid }\overrightarrow{v}\right) +\
^{\shortmid }\overrightarrow{v}\rfloor \ ^{\shortmid }\widehat{\mathbf{D}}_{h%
\mathbf{X}}^{-1}\left( \ ^{\shortmid }\overrightarrow{v}\wedge \ ^{\shortmid
}\widehat{\mathbf{D}}_{h\mathbf{X}}\right) , \\
v\ ^{\shortmid }\widehat{{\mathfrak{R}}} &=&\ ^{\shortmid }\widehat{\mathbf{D%
}}_{v\mathbf{X}}\left( \ ^{\shortmid }\widehat{\mathbf{D}}_{v\mathbf{X}}+\
^{\shortmid }\widehat{\mathbf{D}}_{v\mathbf{X}}^{-1}\left( \ ^{\shortmid }%
\overleftarrow{v}\cdot \right) \ ^{\shortmid }\overleftarrow{v}\right) +\
^{\shortmid }\overleftarrow{v}\rfloor \ ^{\shortmid }\widehat{\mathbf{D}}_{v%
\mathbf{X}}^{-1}\left( \ ^{\shortmid }\overleftarrow{v}\wedge \ ^{\shortmid }%
\widehat{\mathbf{D}}_{v\mathbf{X}}\right) .
\end{eqnarray*}%
Such s-operators induce corresponding non--stretching maps satisfying the
conditions $\ ^{\shortmid }\widehat{\mathbf{D}}_{h\mathbf{Y}}\left(
h\varsigma \right) _{h\mathbf{X}}=0$ and $\ ^{\shortmid }\widehat{\mathbf{D}}%
_{v\mathbf{Y}}\left( v\varsigma \right) _{v\mathbf{X}}=0$.
\end{enumerate}

The canonical recursion d-operator $\ ^{\shortmid }\widehat{{\mathfrak{R}}}%
=(h\ ^{\shortmid }\widehat{{\mathfrak{R}}},v\ ^{\shortmid }\widehat{{%
\mathfrak{R}}})$ is related to respective bi-Hamiltonian structures for
curve flows determined by nonassociative gravity models and generated
solitonic configurations.

Let us consider some examples of $\tau $-families of cosmological solitonic
configurations, for instance with angular anisotropy \ defined by
distributions $\ \wp =\wp (\tau ,r,\vartheta ,t)$ as solutions of a
respective six classes of solitonic 3-d equations 
\begin{eqnarray}
\partial _{rr}^{2}\wp +\epsilon \partial _{t}(\partial _{\vartheta }\wp
+6\wp \partial _{t}\wp +\partial _{^{t t t}}^{3}\wp ) &=&0,\partial
_{rr}^{2}\wp +\epsilon \partial _{\vartheta }(\partial _{t}\wp +\wp \partial
_{\vartheta }\wp +\partial _{\vartheta \vartheta \vartheta }^{3}\wp )=0,
\label{solitdistr} \\
\partial _{\vartheta \vartheta }^{2}\wp +\epsilon \partial _{t}(\partial
_{r}\wp +6\wp \partial _{t}\wp +\partial _{t t t}^{3}\wp ) &=&0,\partial
_{\vartheta \vartheta }^{2}\wp +\epsilon \partial _{r}(\partial _{t}\wp
+6\wp \partial _{r}\wp +\partial _{rrr}^{3}\wp )=0,  \notag \\
\partial _{tt}^{2}\wp +\epsilon \partial _{r}(\partial _{\vartheta }\wp
+6\wp \partial _{r}\wp +\partial _{rrr}^{3}\wp ) &=&0,\ \partial
_{tt}^{2}\wp +\epsilon \partial _{\vartheta }(\partial _{r}\wp +6\wp
\partial _{\vartheta }\wp +\partial _{\vartheta \vartheta \vartheta }^{3}\wp
)=0,  \notag
\end{eqnarray}%
for $\epsilon =\pm 1$. The physical properties of such solutions are well
known from the theory of nonlinear waves and solitonic hierarchies even to
construct solutions in explicit/ general forms is a very difficult task. We
can take any $\wp (\ ^{\shortmid }u)=\ \ ^{\widehat{a}}\wp (\tau
,x^{1},x^{2},t)$ as a parametric, or an exact solution of any equation (\ref%
{solitdistr}), when the abstract label $\widehat{a}$ states the type of the
corresponding 3-d solitonic equation and chosen solution. In a more general
context, we can consider families of such functions for any shell $%
s=1,2,3,4, $ with respective dependencies on momentum like coordinates $%
p_{a_{3}},E$ etc. For simplicity (to understand the nonlinear behaviour), we
may consider only spacetime solitonic cosmological configurations $\ \ ^{%
\widehat{a}}\wp (\tau ,x^{1},x^{2},t)$ and try to model observable solitonic
hierarchic structure formation for certain 4-d effective Lorentz bases.
Nevertheless, we note that the AFCDM allows to construct solutions with
general dependencies on all phase space coordinates of off-diagonal metrics
and generalized connections.

In this work, we distinguish also another types of solitonic hierarchies $\
^{\widehat{a}}\xi (\tau ,x^{i},t)$. On a phase space $\ ^{\shortmid }%
\mathcal{M}$, we write $\ ^{\widehat{a}}\wp (\tau ,x^{1},x^{2},t)$ for
generating, for instance, vacuum solitonic s-metrics. But we label $\ ^{%
\widehat{a}}\xi (\tau ,\ _{s}^{\shortmid }u))$ if certain solitonic
hierarchies are used for defining effective source functionals (\ref%
{cannonsymparamc2a}), 
\begin{equation}
\ ^{\shortmid }\Im _{\ \ \beta _{s}}^{\star \alpha _{s}}=[~_{1}^{\shortmid }%
\mathcal{K}(\ ^{\widehat{a}}\xi )\delta _{i_{1}}^{j_{1}},~_{2}^{\shortmid }%
\mathcal{K}(\ ^{\widehat{a}}\xi )\delta _{b_{2}}^{a_{2}},~_{3}^{\shortmid }%
\mathcal{K}(\ ^{\widehat{a}}\xi )\delta _{a_{3}}^{b_{3}},~_{4}^{\shortmid }%
\mathcal{K}(\ ^{\widehat{a}}\xi )\delta _{a_{4}}^{b_{4}}].
\label{cosmsolitsourc}
\end{equation}%
If such a cosmological s-metric is constructed to define parametric
solutions of gravitational equations (\ref{nonassocaneinst}) and/or (\ref%
{nonassocymgreq1}), we obtain a "nonlinear mix" of solitonic equations
described by s-metrics with functional coefficients depending both on $\ ^{%
\widehat{a}}\wp $ and $\ ^{\widehat{a}}\xi .$ That why, we consider
functionals of type $\ ^{\shortmid }\mathbf{g}(\tau )=\ ^{\shortmid }g[ \ ^{%
\widehat{a}}\wp ,\ ^{\widehat{a}}\xi ]$ for constructing cosmological
solutions of type (\ref{cosmoffds}). The geometric data for such solitonic
hierarchies can be chosen in certain form to model the structure formation/
distributions and $\tau $-parametric evolution of real matter and DM and DE
configurations.

Finally, we emphasize that LC-configurations $\ _{s}^{\shortmid }\nabla $
can be extracted by imposing \ zero torsion conditions, $\ ^{\shortmid }%
\widehat{\mathbf{T}}_{\ \alpha _{s}\beta _{s}}^{\gamma _{s}}=0,$ as we
explain in \cite{partner02,partner04}. Such conditions can be satisfied, in
general, in nonholonomic form.

\end{document}